\documentclass[hyper,11pt,paper]{article}
\usepackage{jheppub}
\usepackage[usenames,dvipsnames,table]{xcolor}
\usepackage{graphicx,amsmath,amssymb,multirow,array,bm,mathrsfs,bbold,bbm}
\usepackage{epsf,amsfonts,slashed}
\usepackage[numbers,sort&compress]{natbib}
\usepackage[bbgreekl]{mathbbol}
\newcommand{\lie}{\pounds}

\newcommand{\super}[1]{\mathbb{#1}}

\newcommand{\cpt}{\hbox{\tiny CPT}}
\newcommand{\CPT}{\ensuremath{\mathcal{CPT}}}

\newcommand{\KMS}{\ensuremath{{K}}}

\newcommand{\btheta}{\bar{\theta}}
\newcommand{\bt}{\bar{\theta}}

\newcommand{\Aa}{\mathcal{A}}
\newcommand{\Gg}{\mathcal{G}}
\newcommand{\bGg}{\overline{\mathcal{G}}}
\newcommand{\Rr}{\mathcal{R}}

\newcommand{\Rho}{\ensuremath{P}}

\newcommand{\M}{\mathcal{M}}
\newcommand{\WV}{\sf{M}}
\newcommand{\G}{\sf{G}}
\newcommand{\Tr}{\hbox{Tr}}
\newcommand{\beq}{\begin{equation}}
\newcommand{\eeq}{\end{equation}}

\newcommand{\Real}{\hbox{Re}}

\newcommand{\be}{\begin{equation}}
\newcommand{\ee}{\end{equation}}
\newcommand{\ba}{\begin{eqnarray}}
\newcommand{\ea}{\end{eqnarray}}

\newcommand{\wt}[1]{\widetilde{#1}}

\title{
Dissipative hydrodynamics in superspace
}

\author[a]{Kristan Jensen,}
\author[b,c]{Natalia Pinzani-Fokeeva,}
\author[b]{and Amos Yarom}
\affiliation[a]{Department of Physics and Astronomy, San Francisco State University, San Francisco, CA 94132, USA}
\affiliation[b]{Department of Physics, Technion, Haifa 32000, Israel}
\affiliation[c]{Department of Mathematics, University of Haifa, Haifa 31905, Israel}

\emailAdd{kristanj@sfsu.edu}
\emailAdd{natascia@physics.technion.ac.il}
\emailAdd{ayarom@physics.technion.ac.il}

\abstract{
We construct a Schwinger-Keldysh effective field theory for relativistic hydrodynamics for charged matter in a thermal background using a superspace formalism. Superspace allows us to efficiently impose the symmetries of the problem and to obtain a simple expression for the effective action. We show that the theory we obtain is compatible with the Kubo-Martin-Schwinger condition, which in turn implies that Green's functions obey the fluctuation-dissipation theorem. Our approach complements and extends existing formulations found in the literature.
}

\begin{document}

\maketitle


\section{Introduction}
\label{S:intro}

The study of fluid mechanics dates back to ancient Greece and the works of Archimedes. Since then, hydrodynamics has undergone countless transformations and modifications before settling into its modern form. Yet, while the dynamics of fluids are prevalent and common, a full understanding of fluid dynamics is still lacking in many respects. 

From a field theoretic viewpoint, relativistic fluid dynamics is a low-energy effective description in terms of constitutive relations for the stress tensor and other conserved currents. The constitutive relations allow one to solve the associated conservation equations and obtain the universal behavior of fully retarded, thermal correlation functions. Up until the works of \cite{Haehl:2015foa,Crossley:2015evo,Haehl:2015uoc}, see also \cite{Haehl:2016pec,Haehl:2016uah}, there was no theory from which one could consistently evaluate symmetric, advanced and other correlation functions associated with the dynamics of the fluid. More simply, fluid dynamics did not follow from an action principle. It was in this sense that hydrodynamics was incomplete. The current treatise merges the somewhat orthogonal constructions of \cite{Crossley:2015evo} and \cite{Haehl:2015foa,Haehl:2015uoc}. We will elaborate on the differences between our formalism and that in the literature when appropriate.

The current work and that of \cite{Haehl:2015foa,Crossley:2015evo,Haehl:2015uoc} do not stand by themselves. A variational principle for dissipationless fluid dynamics was formulated during the last century in the context of general relativity, see e.g., \cite{1911AnP...341..493H} or also \cite{PhysRev.94.1468,Carter}. This variational principle was recently revisited  and recast in modern language in \cite{Dubovsky:2005xd,Andersson:2006nr,Dubovsky:2011sj}. Contemporaneously with these developments, the authors of \cite{Jensen:2012jh,Banerjee:2012iz} argued that hydrodynamics simplifies dramatically in hydrostatic equilibrium. Moreover, in that limit, the constitutive relations can be obtained from a local generating function. While these approaches shed light on the structure of a possible effective action for hydrodynamics  and offer an alternative to some phenomenological approaches~\cite{Bhattacharyya:2013lha,Bhattacharyya:2014bha}, they are lacking in several aspects. Apart from failing to capture dissipation, they do not account for all possible non-dissipative transport phenomena. (For instance, they fail to account for Hall viscosity~\cite{Bhattacharya:2012zx,Haehl:2014zda,Haehl:2015pja}.) Yet another modern approach to obtain an action for hydrodynamics involves adding the effects of stochastic noise~\cite{Kovtun:2014hpa}. Other attempts include \cite{Endlich:2012vt,Andersson:2013jga,Grozdanov:2013dba,Floerchinger:2016gtl}. Most recently, the authors of~\cite{Grozdanov:2013dba,Haehl:2014zda,Haehl:2015pja,Harder:2015nxa} have advocated that the Schwinger-Keldysh formalism is the natural setting to write an effective action for dissipative fluid dynamics. 

The Schwinger-Keldysh formalism \cite{Keldysh:1964ud,Schwinger:1960qe} was developed around the middle of the past century in order to obtain a generating function for connected correlators in a state described by a density matrix $\rho_{-\infty}$ in the far past. Recall that vacuum correlation functions  can be computed by an appropriate variation of the vacuum generating functional with respect to sources, with one such source for each operator. The Schwinger-Keldysh generating functional naturally has two sources associated with each operator. This feature allows one to not only compute the retarded correlation functions in the state $\rho_{-\infty}$, but also partially symmetric and advanced ones.

Let us review the definition and attributes of the Schwinger-Keldysh partition function. We begin with a generic quantum field theory which, in the infinite past, is in a mixed state characterized by a density matrix $\rho_{-\infty}$ (which is not necessarily normalized). The Schwinger-Keldysh partition function is given by
\beq
\label{E:SKdef}
Z[A_1,A_2] \equiv \Tr\left( U_1[A_1] \rho_{-\infty} U_2^{\dagger}[A_2]\right)\,.
\eeq
Here $U_1[A_1]$ is the time-evolution operator, evolving states from the infinite past to the infinite future. It is a functional of external sources which we schematically denote as $A_1$. The time evolution operator $U_2[A_2]$ is similarly defined, and is distinct from $U_1$ via its dependence on $A_2$. 

For a quantum field theory with a Lagrangian description, the Schwinger-Keldysh partition function may be written as a functional integral. If we denote the fundamental fields of the theory by $\phi$ and the action by $S[\phi]$ then, 
\beq
\label{E:SKdef2}
Z[A_1,A_2] = \int [d\phi_1][d\phi_2] \exp\left[ i\left(  S[\phi_1]-S[\phi_2]+\int d^dx \, \left( O_1[\phi_1] A_1-O_2[\phi_2]A_2\right)\right)\right]\,.
\eeq
Here $O$ is the operator conjugate to $A$, and the fields $\phi_1$ and $\phi_2$ satisfy boundary conditions in the infinite past and future. In the past, the boundary conditions depend on $\rho_{-\infty}$. In the future, they are identified, $\lim_{t\to\infty} \phi_1(t) = \lim_{t\to\infty} \phi_2(t)$. We also require that the sources asymptote to the same values in the past and future, $\lim_{t\to\infty} A_1(t) = \lim_{t\to\infty} A_2(t)$ and $\lim_{t\to-\infty} A_1(t)=\lim_{t\to-\infty}A_2(t)$. See e.g. \cite{kamenev2011field,Crossley:2015evo} for a modern discussion. 

Equation \eqref{E:SKdef2} gives an ultraviolet description of the Schwinger-Keldysh partition function for theories with a Lagrangian description. Due to the universality of hydrodynamics it is expected that when the initial state is thermal, i.e. $\rho_{-\infty} = e^{-b H}$ (or its variants with a chemical potential), the infrared behavior of $Z$ will also be universal. More precisely, following the usual logic of Wilsonian effective field theory, we may write
\begin{equation}
	Z[A_1,\,A_2] = \int [d\xi_1][d\xi_2] e^{i S_{eff}[\xi_1,\xi_2; A_1,A_2]}\,,
\end{equation}
when the sources $A_1$ and $A_2$ vary over arbitrarily long scales. Here $S_{eff}[\xi_1,\,\xi_2;A_1,A_2]$ is a low-energy Schwinger-Keldysh effective action with (doubled) infrared degrees of freedom $\xi_1$ and $\xi_2$. We  expect that $S_{eff}$ is universal, and further that it may be viewed as an effective action for hydrodynamics.

In order to obtain an expression for $S_{eff}$ in terms of the infrared degrees of freedom $\xi_1$ and $\xi_2$ one follows the standard path taken when constructing effective actions. Namely, one identifies the infrared degrees of freedom of the theory, the fundamental symmetries associated with their dynamics, and then constructs the most general action compatible with those symmetries. In the context of Schwinger-Keldysh actions for thermal states, this line of research was pioneered in \cite{Crossley:2015evo,Haehl:2015foa,Haehl:2015uoc}.

 Both \cite{Crossley:2015evo} and \cite{Haehl:2015foa,Haehl:2015uoc} have established  that $S_{eff}$ possesses a  nilpotent symmetry transformation reminiscent of supersymmetry.  This is not the first time that such a nilpotent symmetry appears in the context of dissipative dynamics, the canonical example being the Langevin equation, c.f., \cite{PhysRevLett.43.744} or \cite{zinn2002quantum} and references therein. The authors of \cite{Haehl:2015foa,Haehl:2015uoc}  have proposed an a' priori superspace construction which is naturally associated with the formalism developed in \cite{Haehl:2014zda,Haehl:2015pja} in order to capture the symmetries of $S_{eff}$.  The authors of \cite{Crossley:2015evo} have implemented the symmetries associated with $S_{eff}$ in a more direct manner and observed an ``emergent'' superalgebra reminiscent of supersymmetry. In this work we offer a hybrid construction where we build an effective action $S_{eff}$ using a super-Lagrangian from the ground up. 

In section \ref{S:Ingredients} we provide a comprehensive discussion of the symmetries of the Schwinger-Keldysh generating function, the resulting supersymmetry algebra, and how to implement this symmetry in an effective action. In section \ref{S:dof} we elaborate on the low-energy degrees of freedom of the infrared theory.  In section \ref{S:action} we consider a configuration with fixed temperature and velocity, such that the only dynamical field is the chemical potential. In this probe limit we demonstrate that the action principle developed is compatible with known constitutive relations and with the fluctuation-dissipation theorem. While our formalism is similar to that of \cite{Haehl:2015foa,Haehl:2015uoc}, the physical reasoning is comparable to that of \cite{Crossley:2015evo}. Our resulting action differs from that of both groups.  We discuss the differences and similarities of the different approaches in section \ref{S:Discussion} where we also provide an outlook.

While this work was being completed we became aware of~\cite{Liu} by Gao and Liu, which has overlap with this work, and of~\cite{Haehl:2016pec,Haehl:2016uah} which has overlap with some parts of section~\ref{S:Ingredients}.


\section{Symmetries}
\label{S:Ingredients}

The main challenge in constructing any effective theory is to identify its degrees of freedom and it symmetries. In what follows we will study exact symmetries associated with the Schwinger-Keldysh partition function \eqref{E:SKdef} and an initial thermal state. 
Our exposition largely draws on results obtained in the recent work of Crossley, Glorioso and Liu \cite{Crossley:2015evo}, henceforth CGL, and Haehl, Loganayagam and Rangamani \cite{Haehl:2015uoc,Haehl:2015foa} (see also \cite{Haehl:2016pec,Haehl:2016uah}), henceforth HLR.  These two approaches to construct the Schwinger-Keldysh effective action  are similar but not quite the same. In a sense, our work provides a distillation of the superspace formulation of HLR \cite{Haehl:2015uoc,Haehl:2015foa} with the approach of CGL \cite{Crossley:2015evo}. In section \ref{S:Discussion} we discuss similarities and differences between the two approaches and ours.

Recall that the Schwinger-Keldysh generating function for a generic initial state $\rho_{-\infty}$ is given by \eqref{E:SKdef} or by \eqref{E:SKdef2} when a Lagrangian description of the ultraviolet theory is available. In this work we primarily study the case when the initial state is thermal, e.g., $\rho_{-\infty} = \exp(-b H)$.  In this instance the boundary conditions in the past are implemented by an additional segment in the integration contour along imaginary time,
\beq
Z[A_1,A_2] = \int [d\phi_1][d\phi_2][d\phi_E] \exp\Big[ i \big( S[\phi_1,A_1]-S[\phi_2,A_2]\big)\Big]\exp\Big[ - S_E[\phi_E,A_E]\Big]\,,
\eeq
where $S_E$ is the Euclideanized action, and the $A_E$ are the time-independent sources that characterize the initial state. For example, the initial state may be a thermal state on $\mathbb{R}\times \mathbb{S}^{d-1}$, and the radius of the $\mathbb{S}^{d-1}$ would be just such a source. The $A_E$ are related to the sources $A_1$ and $A_2$ in the far past by $ \lim_{t\to-\infty}A_1(t)=\lim_{t\to-\infty}A_2(t)=A_E$. 

Recalling the definition of the partition function $Z[A_1,A_2] = \text{Tr} ( U_1[A_1] \rho_{-\infty} U_2^{\dagger}[A_2])$, we see that variations of the Schwinger-Keldysh generating functional, $W= - i \ln Z$, lead to connected correlation functions of the form
\beq
\label{E:SKcorrelators}
\Tr\left( \frac{\rho_{-\infty}}{\hbox{Tr}\left(\rho_{-\infty}\right)}  \mathcal{\widetilde{T}}\Big( O(\tau_1)\hdots O(\tau_m)\Big) \mathcal{T}\Big( O(t_n)\hdots O(t_1)\Big)\right)\,,
\eeq
where $\mathcal{T}$ and $\mathcal{\widetilde{T}}$ denote the time-ordering and anti-time-ordering  operators respectively, and $O$ is the operator conjugate to $A$. The first string of operators on the right-hand side of \eqref{E:SKcorrelators} comes from the variation of $U_2^{\dagger}$ and the second from the variation of $U_1$.

The chief virtue of the Schwinger-Keldysh formalism is that it computes correlation functions with a wide family of operator orderings. For instance, the symmetric, retarded, and advanced two-point functions of $O$ all follow from variations of $W$,
\begin{align}
\begin{split}
\label{E:Greens}
	G_{sym}(t_1,\,t_2) = \frac{1}{2} \Tr\left(  \frac{\rho_{-\infty}}{\hbox{Tr}\left(\rho_{-\infty}\right)}  \{O(t_1),O(t_2)\}\right)= \frac{\delta^2 W}{\delta A_a(t_1) \delta A_a(t_2)}\Big|_{A_a=A_r=0} \,,\\
	G_{ret}(t_1,\,t_2)	 = i \theta(t_1-t_2) \Tr\left(  \frac{\rho_{-\infty}}{\hbox{Tr}\left(\rho_{-\infty}\right)}  \left[O(t_1),\,O(t_2)\right] \right) =i \frac{\delta^2 W}{\delta A_a(t_1) \delta A_r(t_2)}\Big|_{A_a=A_r=0} \,, \\
	G_{adv}(t_1,\,t_2)	 = -i \theta(t_2-t_1) \Tr\left(  \frac{\rho_{-\infty}}{\hbox{Tr}\left(\rho_{-\infty}\right)}  \left[O(t_1),\,O(t_2)\right] \right)=i \frac{\delta^2 W}{\delta A_r(t_1) \delta A_a(t_2)}\Big|_{A_a=A_r=0}\,,
\end{split}
\end{align}
where we have gone to the so-called $r/a$ basis and defined the average and difference quantities
\beq
\label{E:randa}
	A_r = \frac{1}{2} \left(A_1+A_2\right)\,,
	\qquad
	A_a = A_1-A_2\,,
	\qquad
	O_r = \frac{1}{2}\left(O_1+O_2\right)\,,
	\qquad
	O_a = O_1-O_2\,.
\eeq
In the $r/a$ basis, $r$-type operators are conjugate to $a$-type sources and vice versa,
\beq
\label{E:raBasis}
     \int d^dx\left(O_1A_1-O_2A_2\right)=\int d^dx\left(O_rA_a+O_aA_r\right)\,.
\eeq
For these reasons we will use, e.g.,
\beq
	G_{raa} = \frac{\delta^3 W}{\delta A_a \delta A_r \delta A_r} \Big|_{A_a=A_r=0}\,.
\eeq
In this notation, 
\begin{equation}
\label{E:rasToar}
	G_{ret}= iG_{ra}\,, 
	\qquad
	G_{adv}= iG_{ar}\,,
	\qquad
	G_{sym}= G_{rr}\,.
\end{equation}
We refer the reader to \cite{Crossley:2015evo} for a modern discussion.

We note in passing that equation~\eqref{E:SKcorrelators} makes it clear that not all operator orderings can be obtained from the Schwinger-Keldysh partition function, including the out-of-time-ordered four-point functions which diagnose the onset of chaos~\cite{Maldacena:2015waa,1969JETP...28.1200L}.

After gaining some familiarity with the Schwinger-Keldysh partition function \eqref{E:SKdef} we will, in the remainder of this section, discuss some of its symmetries and the expected infrared degrees of freedom required to describe hydrodynamic behavior. We will focus on four symmetries of the partition function which are independent of the dynamics of the microscopic theory and are generated as a result of the special structure of the Schwinger-Keldysh partition function or, in one instance, are a marked feature of thermal states \cite{Crossley:2015evo,Haehl:2015uoc}. These four symmetries are: 

\begin{enumerate}
	\item \emph{Doubled symmetries}. 
	\label{L:doubled}
	The functional integral representation~\eqref{E:SKdef2}  makes it clear that, in the absence of gravitational anomalies, $Z$ has a doubled reparameterization invariance.  The first weight in the Schwinger-Keldysh functional integral, involving the $1$ fields, can be written with any choice of coordinates, and so can the second. This is true for any initial state. Similarly, if the microscopic theory has a flavor symmetry group $\G$, then  $Z$ is invariant under a doubled flavor gauge invariance, whereby the $1$ and $2$ weights in the Schwinger-Keldysh functional integral may be expressed in different flavor gauges. 

	\item \emph{Topological Schwinger-Keldysh symmetry}. 
	\label{L:SK}
	Consider the Schwinger-Keldysh parition function \eqref{E:SKdef}. If we align the sources of the partition function such that $A_1=A_2$ then unitarity and cyclicity of the trace imply that
	\beq
	\label{E:topologicalLimit}
	Z[A_1=A_2=A] = \Tr\Big( U[A] \rho_{-\infty}U^{\dagger}[A]\Big) = \Tr(\rho_{-\infty})\,.
	\eeq
	A normalized $Z[A_1=A_2=A]$ is independent of the sources $A$. Otherwise, it may depend on the values of the sources in the initial state $A_E$ through $\rho_{-\infty}$.  Going to the $r/a$ basis \eqref{E:randa}, equation \eqref{E:topologicalLimit} implies that when $a$-type sources are set to zero, all variations with respect to the $r$-type sources at times $t>-\infty$ must vanish. Thus, in particular,
	\beq
	\label{E:aTop}
	G_{aa\hdots a} = 0\,.
	\eeq
	We conclude that the Schwinger-Keldysh partition function becomes topological when $A_1=A_2$.
	\item \emph{Reality and positivity}. 
	\label{L:reality}
	As emphasized by~\cite{Crossley:2015evo,Haehl:2015uoc,Haehl:2016pec}, the complex conjugate of $Z$ is given by
	\begin{equation}
	\label{E:realityofZ}
	Z[A_1,A_2]^* = \Tr\Big( U_2[A_2^*]\rho_{-\infty} U_1^{\dagger}[A_1^*]\Big) = Z[A_2^*,A_1^*]\,	
	\end{equation}
	for any Hermitian initial state and complexified sources. In terms of the generating functional $W=-i\ln Z$ the condition~\eqref{E:realityofZ} amounts to
	\beq
	\label{E:reality}
		W[A_1,A_2]^* = - W[A_2^*,A_1^*]\,.
	\eeq
	Equation \eqref{E:reality} is the Schwinger-Keldysh analogue of the usual statement that unitarity implies that the Wilsonian effective action is real, and for this reason we call this a reality condition. However, as~\eqref{E:reality} allows for the effective action $S_{ eff}$  to have an imaginary part, we will also restrict the imaginary part of $S_{ eff}$ for the functional integral to converge.
	\item \emph{KMS symmetry}. 
		\label{L:KMS}
	The partition function possesses an additional symmetry when the initial state is thermal. In the absence of conserved charges, an initial thermal state has the form $\rho_{-\infty}=e^{-b  H}$. This	is the time evolution operator in imaginary time, translating $t \to t - i b$. Thus,
	\beq
	\label{E:KMS}
		Z[A_1(t_1),A_2(t_2)] = \Tr \Big( U_2^{\dagger}[A_2(t_2) ]e^{- b H} U_1[{A}_1(t_1-ib)]\Big)\,,
	\eeq
	which may also be generalized to initial states at nonzero chemical potential. Equation \eqref{E:KMS} leads to the usual statement of the Kubo-Martin-Schwinger (KMS) condition for thermal correlation functions \cite{Kubo:1957mj,PhysRev.115.1342,Haag1967}. Following \cite{Crossley:2015evo},~\eqref{E:KMS} together with CPT invariance leads to a non-local $\mathbb{Z}_2$ symmetry of the partition function given by~\eqref{E:KMSduality}. 
\end{enumerate}

The topological Schwinger-Keldysh symmetry and the reality condition (points \ref{L:SK} and \ref{L:reality}) are a direct result of the definition of the Schwinger-Keldysh partition function \eqref{E:SKdef}. The existence of doubled symmetries (point \ref{L:doubled}) and the KMS symmetry (point \ref{L:KMS}) require a more detailed explanation. In what follows we present an elaborate discussion of the latter.

\subsection{Doubled symmetries}
\label{S:doubled symmetries}

In writing the $r$ and $a$-type sources as in \eqref{E:randa} we have glanced over a subtle point, which we have not seen discussed elsewhere in the literature. In order to construct the $r$ and $a$-type combinations, we need to compare the $1$ and $2$-type operators and sources at the same point. But, in principle, we could use different coordinates $x_1$ and $x_2$ when giving a functional integral description of the time-evolution operators $U_1$ and $U_2$.  Or, to make the issue more severe, suppose that the source we turn on is an external metric, viz. $Z[g_{1\,\mu\nu}(x_1),g_{2\,\rho\lambda}(x_2)]$. In this case, $U_1$ is the time-evolution operator on a spacetime $\M_1$ which differs from the spacetime $\M_2$ on which $U_2$ evolves time. In order to construct the $r$ and $a$-type operators one needs a method by which a point $x_1$ on $\M_1$ can be compared with a point $x_2$ on $\M_2$. 

In order to resolve the issue raised in the previous paragraph, we require that $\M_1$ and $\M_2$ are diffeomorphic to each other; a diffeomorphism from $\M_1$ to $\M_2$ associates a point $x_2$ in $\M_2$ with a point $x_1$ in $\M_1$. This is a global restriction on the sources appearing in the Schwinger-Keldysh partition function. Equivalently, there exists an ``auxiliary spacetime'' $\WV$, which is diffeomorphic to $\M_1$ and $\M_2$, and we can use any diffeomorphism from $\WV$ to $\M_1$ and $\M_2$ to form average and difference combinations on $\WV$. At first sight, it might seem ill-advised to introduce yet another spacetime, especially if we do not have to. However, doing so has the advantage that it allows us to treat the $1$ and $2$ fields on an equal footing, which will prove useful soon. In what follows we will call $\WV$ a ``worldvolume'' and $\M_1$ and $\M_2$ the ``target spaces,'' in analogy with a sigma model. 

Let $\sigma^i$ denote worldvolume coordinates. Then the diffeomorphisms from $\WV$ to $\M_1$ and $\M_2$ are locally represented by maps $x_1^{\mu}(\sigma^i)$ and $x_2^{\mu}(\sigma^i)$, which we can use to pull back the metrics on $\M_1$ and $\M_2$ as
\beq
\label{E:subtlety}
	g_{1\,ij}(\sigma) = g_{1\,\mu\nu}(x_1(\sigma))\partial_{i} x_{1}^{\mu} \partial_j x_{1}^{\nu}\,,
	\qquad
	g_{2\,ij}(\sigma) = g_{2\,\mu\nu}(x_2(\sigma))\partial_{i} x_{2}^{\mu} \partial_j x_{2}^{\nu}\,.
\eeq
This allows us to properly define the average and difference metrics
\beq
\label{E:properra}
	g_{r\,ij}(\sigma) = \frac{1}{2} \left(g_{1\,ij}(\sigma)+g_{2\,ij}(\sigma)\right)\,,
	\qquad
	g_{a\,ij}(\sigma) = g_{1\,ij}(\sigma) - g_{2\,ij}(\sigma)\,.
\eeq
The Schwinger-Keldysh partition function can then be rewritten in terms of $g_{a\,ij}$, $g_{r\,ij}$, and $x_1(\sigma)$ and $x_2(\sigma)$,
\beq
\label{E:ZWV1}
Z[g_{1\,\mu\nu}(x_1),g_{2\,\rho\lambda}(x_2)] = Z[g_{r\,ij}(\sigma),g_{a\, kl}(\sigma);x_{1}(\sigma),x_{2}(\sigma)]\,.
\eeq
The functional variations of $Z$ with respect to $g_{a\,ij}$ and $g_{r\,kl}$ yield correlation functions of operators which could be called the ``average'' and ``difference'' stress-energy tensors $T_r^{ij}$  and $T_a^{kl}$ respectively.

There is a similar story when the microscopic theory has a flavor symmetry group $\G$, in which case we can turn on external gauge fields which couple to the flavor symmetry current. The gauge field $B_{1\,\mu}(x_1)$ on $\M_1$ will generally differ from  $B_{2\,\nu}(x_2)$ on $\M_2$.  Since the external gauge fields $B_1$ and $B_2$ are connections on principal $\G$ bundles over $\M_1$ and $\M_2$, then the analogue of the requirement that $\M_1$ and $\M_2$ are diffeomorphic is that these bundles are isomorphic. Equivalently, the worldvolume $\WV$ too has a principal $\G$ bundle which is isomorphic to those over $\M_1$ and $\M_2$.

In this paper we consider theories with $U(1)$ flavor symmetries, in which case the bundle isomorphism is locally represented by maps $c_1(\sigma)$ and $c_2(\sigma)$. The maps are ``bifundamental'' under target space gauge transformations $\Lambda_{1}$ and $\Lambda_{2}$, as well as under worldvolume gauge transformations $\Lambda$, in the sense that
\beq
\label{E:cTransformation}
	c_{1} \to c_{1} -\Lambda_{1} + \Lambda\,, \qquad
	c_{2} \to c_{2} -\Lambda_{2} + \Lambda\,.
\eeq
The worldvolume transformation law is a consequence of the isomorphisms between the target space and worldvolume bundles. The target space gauge fields pull back to
\beq
\label{E:simplepullback}
	B_{1\, i}(\sigma) = \partial_i x^{\mu}_{1}B_{1\,\mu}(x_{1}(\sigma)) + \partial_i c_{1}\,,
	\qquad
	B_{2\, i}(\sigma) = \partial_i x^{\mu}_{2}B_{2\,\mu}(x_{2}(\sigma)) + \partial_i c_{2}\,.
\eeq
Both $B_i$'s are inert under target space diffeomorphisms and  gauge transformations ($B_{1\,\mu}\to B_{1\,\mu} + \partial_{\mu}\Lambda_{1}$ and $B_{2\,\mu}\to B_{2\,\mu} + \partial_{\mu}\Lambda_{2}$) and transform as connections under worldvolume gauge transformations, $B_{1\,i} \to B_{1\,i} + \partial_i \Lambda$ and $B_{2\,i} \to B_{2\,i} + \partial_i \Lambda$.
 From the two $B_i$'s we define average and difference gauge fields $B_{r\,i}$ and $B_{a\,j}$, such that the partition function can be formally written as
\beq
\label{E:ZWV2}
Z[B_{1\,\mu}(x_1),B_{2\,\nu}(x_2)] = Z[B_{r\,i}(\sigma),B_{a\,j}(\sigma);c_{1}(\sigma),c_{2}(\sigma)]\,.
\eeq
As before, the variation of $Z$ with respect to $B_{a\,i}$ and $B_{r\,j}$ define ``average'' and ``difference'' symmetry currents $J_r^i$ and $J_a^j$ respectively.

Equations \eqref{E:ZWV1} and \eqref{E:ZWV2} immediately imply that the partition function does not depend on the embeddings,
\beq
\label{E:deltaxc}
	\frac{\delta Z}{\delta x_1^{\mu}(\sigma)} = 0\,, \qquad \frac{\delta Z}{\delta x_2^{\nu}(\sigma)} = 0\,,\qquad \frac{\delta Z}{\delta c_1(\sigma)} = 0\,, \qquad \frac{\delta Z}{\delta c_2(\sigma)} = 0\,.
\eeq
It is instructive (and will become useful later) to see how this independence manifests itself in terms of the Ward identities for target space operators.

For simplicity, suppose that the only sources we turn on are external metrics and external $U(1)$ fields. The variation of the Schwinger-Keldysh generating functional $W = - i \ln Z$ may be written as
\begin{align}
\begin{split}
\label{E:deltaW}
\delta W = \int d^dx_1 \sqrt{-g_1}& \left( \frac{1}{2}T_1^{\mu\nu} \delta g_{1\,\mu\nu} + J^{\mu}_1 \delta B_{1\,\mu}\right)
\\
& - \int d^dx_2 \sqrt{-g_2} \left( \frac{1}{2}T_2^{\mu\nu} \delta g_{2\,\mu\nu} + J_2^{\mu} \delta B_{2\,\mu}\right)\,,
\end{split}
\end{align}
where $T_1^{\mu\nu}$ and $T_2^{\mu\nu}$ are the target space stress tensors, and $J_1^{\mu}$ and $J_2^{\mu}$ are the target space $U(1)$ currents.

The reparameterization and $U(1)$ symmetries imply that $W$ is invariant under the combination of infinitesimal target space reparameterizations $\xi_1^{\mu}$ and $\xi_2^{\mu}$, as well as infinitesimal target space $U(1)$ transformations $\Lambda_1$ and $\Lambda_2$. Notating the combined variation as $\delta_{\chi}$, the variation of the external fields under these infinitesimal transformations is
\begin{align}
\begin{split}
\label{E:deltagB}
\delta_{\chi} g_{1\,\mu\nu} & = \pounds_{\xi_1} g_{1\,\mu\nu} = D_{1\,\mu} \xi_{1\,\nu} +D_{1\,\nu}\xi_{1\,\mu}\,,
\\
\delta_{\chi} B_{1\,\mu} & = \pounds_{\xi_1}B_{1\,\mu} +\partial_{\mu}\Lambda_1= - \xi_1^{\nu}G_{1\,\mu\nu} + \partial_{\mu} (\xi_1^{\nu}B_{1\,\nu} + \Lambda_1)\,,
\end{split}
\end{align}
and similarly for the $2$ fields. Here $\pounds_{X}$ is the Lie derivative along $X^{\mu}$, $D_{1\,\mu}$ is the covariant derivative using the Levi-Civita connection constructed from the metric $g_{1\,\mu\nu}$, and $G_{1\,\mu\nu}=\partial_{\mu} B_{1\,\nu} - \partial_{\nu}B_{1\,\mu}$ is the field strength of $B_{1\,\mu}$. Plugging these variations into $\delta W$~\eqref{E:deltaW}, we see that the invariance of $W$ is equivalent to the Ward identities,
\beq
\label{E:doubledWard}
\delta_{\chi}W = 0 \Leftrightarrow \begin{cases} D_{1\,\nu}T_1^{\mu\nu}  = G_1^{\mu}{}_{\nu} J_1^{\nu}- B_1^{\mu} D_{1\,\nu} J_1^{\nu}\,, & D_{1\,\mu} J_1^{\mu} = 0\,,
\\
D_{2\,\nu}T_2^{\mu\nu}  = G_2^{\mu}{}_{\nu}J_2^{\nu}- B_2^{\mu} D_{2\,\nu}J_2^{\nu}\,,& D_{2\,\mu} J_2^{\mu} = 0\,, \end{cases}
\eeq
where in the first line we used $g_{1}^{\mu\nu}$ to raise indices and in the second line we used $g_2^{\mu\nu}$ to do the same.

Now consider expressing the generating functional as a functional of sources pulled back to the worldvolume $\WV$. As the partition function does not explicitly depend on the maps $x_1^{\mu}$, $x_2^{\mu}$, etc., the variation of $W$ may be expressed as
\beq
\label{E:deltaWWV}
\delta W = \int d^d\sigma \left\{ \sqrt{-g_1} \left( \frac{1}{2}T_1^{ij} \delta g_{1\,ij} + J_1^i \delta B_{1\,i} \right) - \sqrt{-g_2}\left( \frac{1}{2}T_2^{ij} \delta g_{2\,ij} + J_2^i \delta B_{2\,i} \right)\right\}\,,
\eeq
where $\sqrt{-g_1}$ and $\sqrt{-g_2}$ are understood to be the measure factors associated with $g_{1\,ij}$ and $g_{2\,ij}$ respectively. We decompose the variations of the worldvolume sources into variations of the target space sources and maps, e.g.
\begin{align}
\begin{split}
\label{E:deltagBWV}
\delta g_{1\,ij}(\sigma) &= \partial_i x_1^{\mu} \partial_j x_1^{\nu}\left(  \delta g_{1\,\mu\nu} + \pounds_{\delta x_1} g_{1\,\mu\nu}\right)\,,
\\
 \delta B_{1\,i}(\sigma) &= \partial_i x_1^{\mu} \left( \delta B_{1\,\mu} + \pounds_{\delta x_1} B_{1\,\mu}+ \partial_{\mu} \delta c_1 \right)\,,
\end{split}
\end{align}
where the Lie derivative is taken with respect to the vector field $\delta x_1^{\mu}(\sigma(x_1))$ on $\M_1$.

Given \eqref{E:deltagBWV} we can match the variation of worldvolume quantities to target space ones. When $\delta x_1^{\mu}=0$, $\delta x_2^{\mu}=0$ etc. comparing~\eqref{E:deltaWWV} with~\eqref{E:deltaW}, we see that the worldvolume stress tensors and currents are related to the target space ones by pushforward, e.g.
\beq
\label{E:fromWVtoTS}
T_1^{\mu\nu} = T_1^{ij} \partial_i x_1^{\mu} \partial_j x_1^{\nu}\,, \qquad J_1^{\mu} = J_1^i \partial_i x_1^{\mu}\,.
\eeq
Allowing for nonzero $\delta x$ and following the same procedure, we find, after integration by parts,
\begin{align}
\begin{split}
\label{E:deltaWWVv2}
\delta W = \int d^d\sigma& \left\{ \sqrt{-g_1} \left( \frac{1}{2}T_1^{\mu\nu} \delta g_{1\,\mu\nu} + J_1^{\mu} \delta B_{1\,\mu} -E_{1\,\mu}  \delta x_1^{\mu} - E_1\delta c_1 \right)\right.
\\
& \quad -\left. \sqrt{-g_2}\left(\frac{1}{2}T_2^{\mu\nu} \delta g_{2\,\mu\nu} + J_2^{\mu} \delta B_{2\,\mu} - E_{2\,\mu} \delta x_2^{\mu} - E_2 \delta c_2 \right)\right\}
\\
& \quad \qquad + (\text{boundary term})\,,
\end{split}
\end{align}
where $E_{1\,\mu}$, $E_{2\,\nu}$, $E_1$, and $E_2$ are the stress tensor and $U(1)$ Ward identities~\eqref{E:doubledWard}, e.g.
\beq
\label{E:wardAgain}
E_{1\,\mu} = D_1^{\nu} T_{1\,\mu\nu} - G_{1\,\mu\nu} J_1^{\nu} + B_{1\,\mu} D_{1\,\nu}J_1^{\nu}\,, \qquad E_1 = D_{1\,\mu} J_1^{\mu}\,.
\eeq
Of course, the Ward identities are satisfied, and so we see that $Z$ does not depend on the maps at all. In equations, 
\beq
\label{E:mapVariation}
\frac{1}{\sqrt{-g_1}}\frac{\delta W}{\delta x_1^{\mu}} = - E_{1\,\mu} =0\,, \qquad \frac{1}{\sqrt{-g_1}}\frac{\delta W}{\delta c_1} = -E_1=0\,,
\eeq
where here the variations are taken when the target space sources are fixed, and there are similar equations for the $2$ fields.

With an eye towards Section \ref{S:dof} (and following earlier works~\cite{Crossley:2015evo,Haehl:2015uoc}) we are unable to resist mentioning that in describing an effective field theory for hydrodynamics the maps $x_1^{\mu}$, $x_2^{\mu}$, $c_1$ and $c_2$ will be promoted to dynamical  fields $X_1^{\mu}$, $X_2^{\mu}$, $C_1$ and $C_2$. Doing so promotes~\eqref{E:mapVariation} (and the corresponding equations for $2$ fields) to field equations, for example
\beq
\frac{1}{\sqrt{-g_1}}\frac{\delta S_{eff}}{\delta X_{1}^{\mu}}=-E_{1\,\mu}=0\,, \qquad\frac{1}{\sqrt{-g_1}} \frac{\delta S_{eff}}{\delta C_{1}}=-E_{1} = 0\,.
\eeq
This guarantees that the effective action is then invariant under the doubled target space symmetries, thereby accounting for the first of the four symmetries of the partition function listed on page \pageref{L:doubled}.

\subsection{Kubo-Martin-Schwinger symmetry}
\label{S:KMS}

Of the four symmetries listed on pages \pageref{L:doubled}-\pageref{L:reality} the KMS symmetry is special in that it is tied to initial states which are in thermal equilibrium and not to generic initial states. Therefore, in order to better understand the KMS symmetry let us take a step back and  briefly review some basics of thermal states from a field theoretic standpoint. We refer the reader to~\cite{Jensen:2013kka} for an extensive discussion. (See also~\cite{Braaten:1995cm} where related techniques were put to work in order to compute the high-temperature free energy of QCD.)

\subsubsection{Thermal states}
\label{SS:Thermal}

Suppose that $\rho_{-\infty}$ is a thermal state characterized by some Boltzmann weight. In order to define this ensemble, we require the existence of a conserved operator $\mathcal{H}$ which generates some notion of time translation so that $\rho_{-\infty} = \exp(-b \mathcal{H})$. In order for $\mathcal{H}$ to be conserved, the sources (e.g., the metric or the flavor field) must be time-translation invariant. Letting $t$ denote time, this means that we can pick a choice of coordinates and flavor gauge so that the background metric and flavor gauge field do not depend on time explicitly and the most general parametrization is 
\begin{align}
\begin{split}
\label{E:staticGauge}
g_{\mu\nu}dx^{\mu}dx^{\nu}& =-e^{2{s}(\vec{x})} (dt + {a}_{\alpha}(\vec{x})dx^{\alpha})^2 + P_{\alpha\beta}(\vec{x})dx^{\alpha}dx^{\beta}\,,
\\
B_{\mu}dx^{\mu} & = B_t(\vec{x})(dt + {a}_{\alpha}(\vec{x})dx^{\alpha}) + B_{\alpha}(\vec{x})dx^{\alpha}\,,
\end{split}
\end{align}
where the parametric length of the Euclidean time circle is $b$. We refer to the choice of coordinates in \eqref{E:staticGauge} as the ``static gauge.''

The expression for the time translation operator $\mathcal{H}$ is given by~\cite{Jensen:2013kka} 
\beq
\label{E:grandPotential0}
	\mathcal{H} = \int d^{d-1}x \sqrt{{\rm det}P_{\alpha\beta}}\,e^{{s}} \big( e^{2{s}}(T^{tt}+T^{t\alpha}{a}_{\alpha} )- J^t  B_t \big)\,.
\eeq
Thus $\exp(-b \mathcal{H})$ is the translation operator in imaginary time, i.e.,
\begin{equation}
	e^{b \mathcal{H}} O(t,\vec{x}) e^{-b \mathcal{H}} = O(t-i b,\vec{x}) \,,
\end{equation}
for any operator $O(t,\vec{x})$. The operator $\mathcal{H}$ is a generalization of what is often called the ``grand canonical potential,'' which for a theory in flat space with constant chemical potential $\mu$ is given by $H-\mu Q$. 

We can now identify an effective, position-dependent temperature $T$ as the inverse length of the thermal circle, a velocity $u^{\mu}$ as the unit norm, future pointing vector with $u^{\alpha}=0$, and a chemical potential $\mu$ proportional to the Polyakov loop,
\beq
	T = \frac{e^{-{s}}}{b}\,, \qquad u^{\mu}\partial_{\mu} = {e^{-{s}}} \partial_t\,, \qquad \mu = e^{-{s}}B_t\,.
\eeq

All of these statements above may be phrased covariantly.  The time-translation invariance of the metric and flavor gauge field may be captured by a Killing vector field $\beta^{\mu}$ and a gauge parameter $\Lambda_{\beta}$. We denote their combined action on the metric and gauge field by $\delta_{\beta}$:
\begin{align}
\begin{split}
\label{E:deltabeta}
\delta_{\beta} g_{\mu\nu} & = \lie_{\beta}g_{\mu\nu} = \beta^{\rho}\partial_{\rho}g_{\mu\nu} + g_{\mu\rho}\partial_{\nu}\beta^{\rho} + g_{\nu\rho} \partial_{\mu}\beta^{\rho} = 0\,,
\\
\delta_{\beta} B_{\mu} & = \lie_{\beta} B_{\mu} + \partial_{\mu}\Lambda_{\beta}= \beta^{\nu}\partial_{\nu}B_{\mu}+B_{\nu}\partial_{\mu}\beta^{\nu} + \partial_{\mu}\Lambda_{\beta} = 0\,,
\end{split}
\end{align}
where $\lie_{\beta}$ is the Lie derivative along $\beta^{\mu}$. The time-translation operator $\mathcal{H}$ is now given by
\beq
\label{E:grandPotential}
	b \mathcal{H} = - \int dV_{\mu} \Big( T^{\mu\nu} \beta_{\nu} + J^{\mu}  (\beta^{\nu}B_{\nu}+\Lambda_{\beta})\Big)\,,
\eeq
where $dV_{\mu}$ is the volume form on a constant-time slice. The covariant version of the statement that $\exp(-b \mathcal{H})$ is a translation operator in imaginary time is that it acts on fields as
\beq
\label{E:timeevolution}
	\exp(b \mathcal{H})O(x^{\mu})\exp(-b \mathcal{H}) = e^{-i \delta_{\beta}}O(x^{\mu}) \,.
\eeq
The temperature, velocity, and chemical potential are given by
\beq
	T = \frac{1}{\sqrt{-\beta^2}}\,, \qquad u^{\mu} = \frac{\beta^{\mu}}{\sqrt{-\beta^2}}\,, \qquad \frac{\mu}{T} = \beta^{\mu}B_{\mu} + \Lambda_{\beta}\,.
\eeq
One can go from the covariant gauge to the static gauge by setting $\beta^{\mu}\partial_{\mu} = b \partial_t$ and $\Lambda_{\beta}=0$.

Let us now turn back to the Schwinger-Keldysh partition function. In what follows, we posit that $\beta$ and $\Lambda_{\beta}$ are thermodynamic parameters of the initial thermal state retained in the Schwinger-Keldysh effective action as fixed, non-dynamical data  and that they live on the worldvolume $\WV$. Moreover, we will implement worldvolume reparametrization and $U(1)$ symmetries in the effective action. In what follows we will equip $\Lambda_{\beta}$ with a transformation law
\beq
\label{E:Lbtransformation}
\delta_{\Lambda} \Lambda_{\beta} = - \beta^{i}\partial_{i}\Lambda\,,
\eeq
and also let $\Lambda_{\beta}$ and $\beta^i$ transform as a scalar and vector respectively under worldvolume diffeomorphisms. With these transformation laws the action we construct will be invariant under worldvolume reparameterizations and $U(1)$ symmetries. It should now be clear that, for example, the  chemical potential $\mu/T=\beta^iB_i+\Lambda_{\beta}$ defined on $\WV$ is invariant under  worldvolume gauge transformations. Note that in static gauge, time-independent gauge transformations are residual symmetries under which $B_i$ varies as $\delta_{\Lambda}B_i=\partial_i\Lambda(\vec{\sigma})$, which is compatible with the findings of \cite{Crossley:2015evo}. 

In the far past there is only one $\beta$ and one $\Lambda_{\beta}$ characterizing both target spaces $\mathcal{M}_1$ and $\mathcal{M}_2$ and the worldvolume manifold $\WV$, which all coincide. Once $A_1$ and $A_2$ are not aligned, we make the convenient choice that $\beta^i(\sigma)$  and $\Lambda_{\beta}(\sigma)$ lie in $\WV$ and their action on fields and sources is denoted by $\delta_{\beta}$. The pushforward of $\beta^i$ and $\Lambda_{\beta}$ to the target spaces is given by 
\begin{align}
\begin{split}
\beta^{\mu}_1(x_1)&=\beta^i(\sigma(x_1))\partial_ix_1^{\mu}\,,\hspace{1.19in} \beta^{\mu}_2(x_2)=\beta^i(\sigma(x_2))\partial_ix_2^{\mu}\,,
\\
\Lambda_{\beta\,1}(x_1)&=\Lambda_{\beta}(\sigma(x_1))+\beta^{\mu}_1\partial_{\mu}c_1(\sigma(x_1))\,,\qquad \Lambda_{\beta\,2}(x_2)=\Lambda_{\beta}(\sigma(x_2))+\beta^{\mu}_2\partial_{\mu}c_2(\sigma(x_2))\,.
\end{split}
\end{align}

With $\beta$ and $\Lambda_{\beta}$ naturally residing on the worldvolume, it is also convenient to choose the time evolution operators $U_1$ and $U_2$ to act on the worldvolume. We construct $U_1$ and $U_2$ on $\WV$ from the operators $\mathcal{H}_1$ and $\mathcal{H}_2$ as in~\eqref{E:grandPotential}, using the fixed initial state data $(\beta^i(\sigma),\Lambda_{\beta}(\sigma))$ and the target space stress tensors and currents pulled back to $\WV$. 

\subsubsection{CPT invariance and KMS symmetry }

 With a covariant formulation of thermal equilibrium at hand, we now carefully go over the steps leading to \eqref{E:KMS} allowing for non zero chemical potential. We begin by reprising the definition of the Schwinger-Keldysh partition function,
\beq
\label{E:SKreminder1}
	Z[A_1,A_2;\beta,\Lambda_{\beta}] = \hbox{Tr} \left(U_1[A_1] e^{-b \mathcal{H}} U_2^{\dagger}[A_2] \right)\,.
\eeq
In \eqref{E:SKreminder1} we have added the dependence of $Z$ on the initial state parameters $\beta$ and $\Lambda_{\beta}$. We will often use such notation when there is an ambiguity regarding the parameters of the initial state. Otherwise we will omit the dependence on the initial thermodynamic parameters for brevity as we have done until now. 

The KMS transformation we advertised on page \pageref{L:KMS} follows from the fact that $\exp(-b \mathcal{H}) $ is the time-evolution operator in imaginary time. In what follows we denote the worldvolume time coordinate by $\sigma^0=\tau$. To wit, in static gauge we find
\beq
	e^{b \mathcal{H}}\left( \mathcal{T}e^{i \int d^d\sigma \, H[A(\sigma),\,O(\sigma)]}\right)e^{-b\mathcal{H}}=\mathcal{T}e^{i \int d^d\sigma \, H[A(\sigma),\, O(\tau-i b,\vec{\sigma})]} = \mathcal{T}e^{i \int d^d\sigma \, H[A(\tau+i b,\vec{\sigma}),\,O(\sigma)]}\,, 
\eeq
where $H$ is the Hamiltonian density, i.e. 
\beq
\label{E:Utransformation}
	U[A (\tau )] \exp(-b\mathcal{H}) = \exp(-b \mathcal{H}) U [A (\tau +i b)] = e^{-b \mathcal{H}} U[e^{i \delta_{\beta}} A(\tau)]\,.
\eeq
Covariantly,
\begin{align}
\begin{split}
\label{E:KMSreprise}
	Z[A_1,\,A_2;\,\beta,\,\Lambda_{\beta}] &= \hbox{Tr} \left(U_1[A_1] e^{-b \mathcal{H}}U_2^{\dagger}[A_2] \right)  \\
		& = \hbox{Tr} \left(U_2^{\dagger}[A_2] e^{-b \mathcal{H}} U_1[e^{i\delta_{\beta}} A_1] \right)= \Tr\left( U_2^{\dagger}[e^{-i\delta_{\beta}}A_2]e^{-b \mathcal{H}} U_1[A_1]\right)\,,
\end{split} 
\end{align}
and we have used the cyclicity of the trace. As emphasized by CGL \cite{Crossley:2015evo} the right-hand side of \eqref{E:KMSreprise} does not correspond to a Schwinger-Keldysh partition function. Rather, using CPT \, one can relate the right-hand side of \eqref{E:KMSreprise} to the CPT-transformed partition function, as we now review.

In what follows we define a(n antilinear) CPT transformation on general spacetimes, which we denote by $\CPT$,~such that $\CPT^2=1$. It is defined as a suitable action on fields as well as a geometric part which acts on spacetime.  If we consider the time direction on our manifold $\WV$ as being fibered over some base manifold $\WV_s$ then $\CPT$ acts by inverting time $\tau \to -\tau$ as well as by an orientation-reversing transformation on $\WV_s$. If $\CPT$ acts on coordinates $\sigma$, we denote the resulting combined coordinate transformation by $\vartheta \sigma$. 

It is an old result in axiomatic perturbative quantum field theory that Lorentz invariance and locality imply invariance under CPT~for a broad class of theories. We are not aware of a general proof for quantum field theory on more general spacetimes, in particular those on which we may define thermal states. In this work we assume that any ``healthy'' quantum field theory on a spacetime of the sort we considered above is invariant under \CPT. This implies that under $\CPT$ a bosonic operator $O(\sigma)$ is transformed as
\beq
(\CPT)O(\sigma)(\CPT) = \eta_O O^*(\vartheta \sigma) \equiv \Theta O(\sigma) \,,
\eeq
where $\eta_O = \pm 1$ is the \CPT-eigenvalue of $O$. Note that in hydrodynamics one usually considers only bosonic operators (see however e.g. \cite{Grassi:2011wt} for discussions of hydrodynamics of supersymmetric material). This will somewhat simplify future expressions. 

Using that \CPT \,is anti-unitary it then follows that
\begin{align}
\begin{split}
\label{E:CPTderivation}
	(\CPT) \mathcal{T} e^{i \int d^d\sigma \, H[A(\sigma),\,O(\sigma)]}(\CPT) & =\wt{\mathcal{T}} e^{-i \int d^d\sigma \, H[A(\sigma)\, ,\,\eta_O O(\vartheta \sigma)]} \\
		&=\wt{\mathcal{T}} e^{-i \int d^d\sigma \, H[\eta_O A(\vartheta\sigma) ,\, O(\sigma)]} \\
		&=\wt{\mathcal{T}} e^{-i \int d^d\sigma \, H[\Theta A(\sigma) ),\, O(\sigma)]}\,,
\end{split}
\end{align}
where $\wt{\mathcal{T}}$ is the  anti-time-ordering operator and in the last equality we have defined the action of $\Theta$ on a (real) source $A$ conjugate to an operator $O$,
\begin{equation}
\label{E:defThetaA}
	\Theta A(\sigma) = \eta_A A(\vartheta \sigma)\,,
\end{equation}
where $\eta_A = \eta_O$. Thus,
\beq
	(\CPT) U[A] (\CPT) = U^{\dagger}[\Theta A]\,.
\eeq
In obtaining \eqref{E:CPTderivation} we have assumed a \CPT~invariant theory  so that $H$ is invariant under a \CPT~transformation of $O$ combined with a spurionic transformation of $A$. We emphasize that \CPT~acts only on operators. The source $A$ does not transform under it. Instead, the effect of \CPT \, on the time-evolution operator is equivalent to the combination of Hermitian conjugation and replacing the source $A(\sigma)$ with $\Theta A(\sigma)$.

Apart from the operators in the theory, the initial state $\rho_{-\infty}$ also transforms under \CPT. With a slight abuse of notation we define
\begin{align}
\begin{split}
\label{E:betaCPT}
  (\CPT)\beta^i(\sigma)(\CPT) &= \eta_{\beta^i} \beta^i( \vartheta \sigma) = \Theta \beta^i(\sigma)\,, 
  \qquad  \hbox{no sum over $i$,}
  \\
  (\CPT)\Lambda_{\beta}(\sigma)(\CPT) &= -\Lambda_{\beta}( \vartheta \sigma) = \Theta \Lambda_{\beta}(\sigma)  
  \,,
\end{split}
\end{align}
where $\eta_{\beta^i}$ is the eigenvalue of the $i$'th component of $\beta^i$ under \CPT. Recall that $\beta^i(\sigma)$ specifies integral curves along the local time direction,
\begin{equation}
	\frac{\partial \sigma^i(\lambda)}{\partial \lambda} = \beta^i(\sigma(\lambda))\,.
\end{equation}
Time reversal flips both $\lambda$ and $\sigma^0=\tau$, and parity flips one of the spatial coordinates, say $\sigma^1$. With these conventions, $\eta_{\beta^0}=1$, $\eta_{\beta^1}=1$ and $\eta_{\beta^i}=-1$ for $i\geq 2$. In even spacetime dimensions one often refers to parity as a combination of an inversion of one of the space coordinates and a rotation of the others. In those conventions we would have $\eta_{\beta^i}=1$ for all $i$. The gauge parameter $\Lambda_{\beta}$ has eigenvalue $-1$ under \CPT~because it changes sign under charge conjugation.

Focusing our attention on a thermal initial state, $\rho_{-\infty} = e^{-b \mathcal{H}}$, we define $\CPT \rho_{-\infty} \CPT = e^{-b \mathcal{H}^{\cpt}}$
and $\delta_{\beta}^{\cpt}$ via
\begin{equation}
\label{E:Thetadeltacommutator}
	\Theta \left( e^{-i\delta_{\beta} } O(\sigma) \right) = 
	e^{i \delta^{\cpt}_{\beta} } \Theta O(\sigma)\,,
\end{equation}
where we remind the reader that $\Theta$ is antilinear. The equivalent of \eqref{E:timeevolution} is 
\beq
	\exp (b \mathcal{H}^{\cpt}) O(\sigma) \exp(-b\mathcal{H}^{\cpt}) = e^{i\delta_{\beta}^{\cpt}}O(\sigma)\,.
\eeq

Let us use the above definitions to bring the KMS-transformed partition function \eqref{E:KMSreprise} to canonical form. A microscopic \CPT~symmetry takes the Schwinger-Keldysh partition function to one in which a future state is evolved backwards in time,
\beq
\label{E:cptZ}
	Z[A_1,A_2;\,\beta,\,\Lambda_{\beta}] = \Tr\Big( U_1[A_1] e^{-b \mathcal{H}} U_2^{\dagger}[A_2]\Big) = \Tr\Big( U_1^{\dagger}[\Theta A_1] e^{-b \mathcal{H}^{\cpt}}  U_2[\Theta A_2]\Big)^* \,,
\eeq
where in the last equality we have used the anti-cyclicity property of the trace of a product of antilinear operators. Using 
\beq
\label{E:Utransformation2}
	\exp\left( b \mathcal{H}^{\cpt} \right) U[A] \exp \left( -b \mathcal{H}^{\cpt} \right) = U[e^{-i\delta_{\beta}^{\cpt}} A ]\,,
\eeq
as in~\eqref{E:Utransformation}, and following the same logic that led us to~\eqref{E:KMSreprise}, we find that~\eqref{E:cptZ} implies 
\begin{align}
\begin{split}
\label{E:KMSv2}
	Z[A_1,A_2;\beta,\Lambda_{\beta}] &= \Tr \left( U_2[e^{i\delta_{\beta}^{\cpt}} \Theta A_2] e^{-b \mathcal{H}^{\cpt}} U_1^{\dagger}[\Theta A_1] \right)^* \\
	& = Z[e^{i\delta^{\cpt}_{\beta} }\Theta A_2,\,\Theta A_1;\,\Theta \beta, \Theta \Lambda_{\beta}]^*\,.
\end{split}
\end{align}
A functional integral proof of \eqref{E:KMSv2} can be found in \cite{Jensen:2018hse}.

We find it useful to implement the reality condition~\eqref{E:realityofZ}
\begin{equation*}
	Z[A_1,A_2]^* = Z[A_2^*,A_1^*]\,,
\end{equation*}
to reexpress~\eqref{E:KMSv2} as
\begin{align}
\begin{split}
\label{E:KMSduality}
	Z[A_1,\,A_2;\,\beta,\,\Lambda_{\beta}] &= Z [ \left(\Theta A_1\right)^*, \left(e^{i \delta^{\cpt}_{\beta}} \Theta A_2\right)^* ;\, \Theta \beta,\, \Theta \Lambda_{\beta}] \\
	&= Z [ \Theta^* A_1, \Theta^* \left(e^{-i \delta_{\beta}} A_2\right);\,\Theta  \beta,\,\Theta  \Lambda_{\beta}] 
	\,,
\end{split}
\end{align}
were in the last line we have defined $\Theta^*$ as a \CPT~transformation followed by complex conjugation. We refer to \eqref{E:KMSduality} as the KMS symmetry and it will be crucial in what follows. In static gauge, equation \eqref{E:KMSduality} becomes
\begin{equation*}
	Z[A_1(\tau_1),A_2(\tau_2)] = Z [\eta_A A_1(-\tau_1), \eta_A A_2 (-\tau_2 - ib)]\,,
\end{equation*}
which is identical to the KMS condition discussed in CGL \cite{Crossley:2015evo}.\footnote{CGL derived a similar condition in~\cite{Crossley:2015evo}, using $\mathcal{P}\mathcal{T}$ rather than \CPT.}

In \eqref{E:KMSduality} it is clear that the KMS symmetry is a $\mathbb{Z}_2$ transformation: acting with it twice brings us back to the original Schwinger-Keldysh partition function,
\begin{align}
\begin{split}
\label{E:realZ2}
	Z[A_1,A_2;\beta,\Lambda_{\beta}] &= Z[A_1, \Theta^*\left(e^{i \delta^{\cpt}_{\beta}} \Theta^* \left( e^{-i\delta_{\beta}}  A_2 \right) \right);\beta,\Lambda_{\beta}] \\
	&= Z[A_1,A_2;\beta,\Lambda_{\beta}]\,.
\end{split}
\end{align}
This $\mathbb{Z}_2$ transformation is both unitary and non-local. It is unitary because it involves both a CPT~flip and complex conjugation. It is non-local in that it shifts the insertion point of operators by a finite distance in imaginary time.

As a sanity check we note that the microscopic Schwinger-Keldysh action,
\begin{equation*}
S_{SK}  = S[\phi_1,A_1] - S[\phi_2,A_2]\,,
\end{equation*}
is consistent with the KMS symmetry. Because the full KMS transformation is unitary, the action must be invariant upon replacing $A_1$ with its \CPT-conjugate and $A_2$ with its \CPT-conjugate as well as a translation in imaginary time. The microscopic action is clearly invariant under this transformation provided we equip the dynamical fields $\phi$ with the same transformations as the sources $A$ and the action $S$ is \CPT-invariant. 

Before ending this section we note that the KMS symmetry leads to an additional topological sector of the theory. This observation will become important once we attempt to implement the symmetries in an effective action in section \ref{S:Implementation}.

Recall that the topological Schwinger-Keldysh symmetry followed from the observation that when the sources are aligned, $A_1 = A_2=A$, the Schwinger-Keldysh partition function reduces to
\begin{equation*}
	Z[A,A] = \text{Tr}\left( U[A]\rho_0 U^{\dagger}[A]\right) = \text{Tr}(\rho_{-\infty})\,,
\end{equation*}
which is independent of the common source $A$. Consequently correlation functions of the conjugate operators vanish. Using that the variation of $W = -i \ln Z$ in terms of $1$ and $2$ fields is given by
\beq
	\label{E:deltaWagain}
	\delta W = \int d^d\sigma \left( O_1\delta A_1 - O_2\delta A_2\right)\,,
\eeq
and plugging in $\delta A_1 = \delta A_2 = \delta A$, we found that $A$ was conjugate to the difference operator, $O_a = O_1 - O_2$. More simply, $Z$ has a topological limit for any initial state $\rho_{-\infty}$.

The KMS symmetry~\eqref{E:KMSduality} implies the existence of a different topological limit when the initial state is thermal, $\rho_{-\infty} = e^{-b\mathcal{H}}$. When $A_1 = A$ and $A_2 = e^{i\delta_{\beta}}A$,~\eqref{E:KMSduality} implies
\begin{align}
\begin{split}
\label{E:topologicalKMS}
	Z[A,e^{i\delta_{\beta}}A;\beta,\Lambda_{\beta}] &= 
	Z[\Theta^*A,\Theta^*A;\Theta\beta,\Theta\Lambda_{\beta}] =
	\text{Tr}\left( U[\Theta^*A]e^{-b \mathcal{H}^{\cpt}}U^{\dagger}[\Theta^* A]\right) 
	\\
	&= \text{Tr}\left( e^{-b \mathcal{H}^{\cpt}}\right)\,,
\end{split}
\end{align}
which is independent of $A$. As above, correlation functions of the conjugate operators vanish. Plugging in $\delta A_1 = \delta A$ and $\delta A_2 = e^{i\delta_{\beta}}\delta A$ into the variation of $W$ in~\eqref{E:deltaWagain}, we see that the operator conjugate to $A$ is what we term the $\tilde{a}$-type operator, given by $\wt{O}_a = O_1 - e^{-i\delta_{\beta}}O_2$. The $\tilde{a}$-type operators comprise an additional topological sector and their correlation functions with each other must vanish. We refer to this property as the topological KMS symmetry.

There is an analogue of the $r/a$ basis which will be useful in what follows. We define the $\tilde{r}/\tilde{a}$ basis as
\beq\label{E:defOaOr}
	\wt{O}_r = \frac{O_1 + e^{-i\delta_{\beta}}O_2}{2}\,,
	\qquad
	\wt{O}_a = O_1 - e^{-i\delta_{\beta}}O_2\,,
\eeq
and similarly define $\tilde{r}$ and $\tilde{a}$-type sources. In static gauge we have
\beq
	\wt{O}_r(\sigma) = \frac{O_1(\sigma) + O_2(\tau-i b,\vec{\sigma})}{2}\,,
	\qquad
	\wt{O}_a(\sigma) = O_1(\sigma) - O_2(\tau-i b,\vec{\sigma})\,.
\eeq
In terms of the tilde'd combinations,~\eqref{E:deltaWagain} may be rewritten as
\beq
	\delta W = \int d^d\sigma\, \left( \wt{O}_r\delta \wt{A}_a + \wt{O}_a \delta \wt{A}_r\right)\,.
\eeq
In particular, the $\wt{A}_r = (A_1+e^{-i\delta_{\beta}}A_2)/2$ sources are conjugate to the $\wt{O}_a$'s. This is consistent with our discussion above: setting $A_1 = A$ and $A_2 = e^{i\delta_{\beta}}A$ sets $\wt{A}_a = 0$ and $\wt{A}_r=A$. 

\section{Dynamical degrees of freedom and an implementation of the symmetries}
\label{S:Implementation}

In order to construct a Schwinger-Keldysh effective action we need to identify its symmetries and degrees of freedom. In the previous section we have discussed the symmetries required by the Schwinger-Keldysh generating function. In what follows we will discuss how these symmetries may be implemented on the dynamical degrees of freedom which enter into the effective action. 

\subsection{Dynamical degrees of freedom and doubled symmetries}

Following~\cite{Crossley:2015evo,Haehl:2015uoc}, we make the ansatz that at low energies the maps $x_1^{\mu},\, c_1$, etc. are promoted to dynamical fields, which we denote as $X_1^{\mu}(\sigma)$, $C_1(\sigma)$, and so on. We consider systems for which these are the only light degrees of freedom. We write down  effective actions $S_{ eff}$ on the worldvolume, and impose that the fields $X_1^{\mu}(\sigma)$, $C_1(\sigma)$, etc., only appear in the action through the pullbacks of the target space sources to the worldvolume. We also demand that the effective action is invariant under worldvolume reparameterizations and $U(1)$ transformations.

More precisely, we define
\begin{align}
\begin{split}
\label{E:defBigi}
	B_{1\,i} &= \partial_i X_1^{\mu} B_{1\,\mu}(X_1)+\partial_i C_1\,,
	\qquad
	B_{2\,i} = \partial_i X_2^{\mu} B_{2\,\mu}(X_2)+\partial_i C_2\,,
	\\
	g_{1\,ij} & = \partial_i X_1^{\mu} \partial_j X_1^{\nu} g_{1\,\mu\nu}(X_1)\,,
	\hspace{.41in}
	g_{2\,ij} = \partial_i X_2^{\mu} \partial_j X_2^{\nu} g_{1\,\mu\nu}(X_2)\,,
\end{split}
\end{align}
so that the action depends on the $X$'s and $C$'s via
\beq
	S_{eff} = \int d^d\sigma \,L_{eff}(g_{1\,ij},\,g_{2\,ij},\,B_{1\,i},\,B_{2\,i},D_i;\,\beta^i,\,\Lambda_{\beta})
\eeq
where $\beta^i$ and $\Lambda_{\beta}$ are parameters of the the initial thermal state as discussed in section \ref{SS:Thermal}.

The reason for this ansatz is the following. In hydrodynamics, one enforces the conservation of the stress tensor $T^{\mu\nu}$ and $U(1)$ current $J^{\mu}$ as equations of motion. In the Schwinger-Keldysh setting, the Ward identities are doubled, as we found in~\eqref{E:doubledWard}. As we will see shortly, choosing to promote the maps to dynamical fields has the desirable property that the equations of motion for the dynamical fields (which are promoted to operator identities in the quantum theory) are precisely  the doubled Ward identities. The Ward identities for the $r$-type stress tensor and $U(1)$ current (with aligned sources) lead to the hydrodynamic equations. In this way effective actions for hydrodynamics are (doubled) sigma models.

To obtain the desired Ward identities consider, as an example, the target space $U(1)$ current $J_{1}^{\mu}$, which is obtained by varying the generating functional with respect to the $U(1)$ source $B_{1\,\mu}$, 
\beq
\label{E:defJ1}
J_1^{\mu} = \frac{\delta S_{eff}}{\delta B_{1\mu}}\,.
\eeq
Using that the effective action only depends on $B_{1\mu}$ through the pullback $B_{1i}$, we write a more general variation of $S_{eff}$ as
\beq
\delta S_{eff} = \int d^d\sigma \,J_1^i \delta B_{1i} = \int d^d\sigma  \,J_1^i  \left(\partial_i X_1^{\mu} \delta B_{1\mu}  + \partial_i \delta C_1\right) = \int d^d\sigma \left( (J_1^i \partial_iX_1^{\mu})\delta B_{1\mu} - (\partial_i J_1^i) \delta C_1\right)\,.
\eeq
Comparing with~\eqref{E:defJ1}, we see that
\beq
J_1^{\mu} = J_1^i\partial_i X_1^{\mu}\,,
\eeq
and then the $C_1$ equation of motion is simply that this current is conserved,
\beq
\frac{\delta S_{eff}}{\delta C_1} = - \partial_{\mu}J_1^{\mu}\,.
\eeq

An analysis similar to the one derived above
shows that the dynamical equations for the $X$'s are identical to conservation of the energy momentum tensor in each of the target spaces. However, as we will see shortly, it is difficult to reconcile the doubled diffeomorphism invariance of the generating function together with the Schwinger-Keldysh topological symmetry. Thus, in most of what follows, we will work in a probe limit, where the target space metrics are identical and are given by the Minkowski metric, and the mappings $X_1^{\mu}$ and $X_2^{\mu}$ are non-dynamical and reduce to the trivial map,
\begin{equation}
\label{E:probe} 
	g_{1\,\mu\nu}=g_{2\,\mu\nu} = \eta_{\mu\nu}
	\qquad
	X_1^{\mu} = X_2^{\mu} = \delta_i^{\mu} \sigma^i\,.
\end{equation}
We will discuss this some more in Subsection~\ref{S:dof} and in the Discussion when comparing our work to others, and extend it in a future publication.

\subsection{Topological Schwinger-Keldysh symmetry}
\label{S:SK}

In the second entry of our list of symmetries on page \pageref{L:SK} we noted that every Schwinger-Keldysh partition function has a {topological limit} when the sources are aligned, $A_1 = A_2$, viz.,
\begin{equation}
\label{E:aTopagain}
	G_{aa\ldots a}=0.
\end{equation}
Another way of stating this result is that in any Schwinger-Keldysh theory the $a$-type operators $O_1-O_2$ form a topological sector.

In what follows we would like to provide a construction which will ensure that this topological sector remains intact in the Wilsonian effective theory. We will start our discussion in section \ref{SS:topological} with a lightning review of Witten-type topological theories and their manifestation in superspace. In Section \ref{SS:SDTQFT} we will discuss how to deform such topological theories in order to capture the non-topological nature of the Schwinger-Keldysh path integral whenever $A_1 \neq A_2$. Finally, in Section \ref{SS:applicationtoSK} we will see how to implement the topological symmetry using a superspace formalism. Our description leans on HLR \cite{Haehl:2015foa,Haehl:2015uoc} and textbook material~\cite{zinn2002quantum,kamenev2011field}. See also the very recent \cite{Haehl:2016pec} which has some overlap with the current section. 

\subsubsection{Cohomological quantum field theories}
\label{SS:topological}

Recall that topological quantum field theories are often defined as quantum field theories in which expectation values of physical operators are independent of the metric,
\begin{equation}
\label{E:CanonTQFT}
	\frac{\delta}{\delta g_{\mu\nu}} \langle O_{i_1} \ldots O_{i_n} \rangle = 0 \,.
\end{equation}
It is common to classify such theories into one of two categories. The first are referred to as Witten (or cohomological)-type quantum field theories \cite{Witten:1988ze,Witten:1990bs}. The other category includes Schwarz (or quantum) topological field theories \cite{Schwarz:1978cn}. An example of Schwarz-type theories is Chern-Simons theory. The Witten-type theories have the following properties. 
\begin{enumerate}
\item
	There exists a Grassmannian operator $Q$ with $Q^2=0$ whose action we represent as $\delta_Q$.
\item
	Physical operators and the action itself are $Q$-closed, i.e., $\delta_Q S = 0$. (So $Q$ can be thought of as a scalar supercharge.)
\item
	The stress tensor is $Q$-exact,
	\begin{equation}
		T^{\mu\nu} = \delta_Q V^{\mu\nu} \,.
	\end{equation}
	Note that $V^{\mu\nu}$ is a {ghost}: it has odd Grassman-parity while carrying integer spin.
\end{enumerate}

The properties described above ensure that the partition function is independent of the metric. When a theory has a functional integral description with action $S[\phi]$, the variation of the partition function with respect to the metric is
\begin{align}
\begin{split}
	\delta_g Z &= \int [d\phi] \delta_g e^{-S[\phi]} \\
	&=\int [d\phi]\int d^dx\sqrt{-g} \left(\frac{1}{2}\delta g_{\mu\nu} \delta_Q V^{\mu\nu}  \right) e^{-S[\phi]} \\
	&=\int [d\phi]\delta_Q \left( \int d^dx\sqrt{-g}\,  \frac{1}{2}\delta g_{\mu\nu} V^{\mu\nu} e^{-S[\phi]} \right)\\
	&=0\,,
\end{split}
\end{align}
where the last equality follows from integration by parts in field space and assuming a $Q$-invariant measure. Note that, being an external source and not a dynamical field, $g_{\mu\nu}$ is inert under $\delta_Q$. A similar argument shows that the correlation functions of any $Q$-closed operator do not depend on the metric, and that correlation functions of $Q$-exact operators must vanish. In particular
\beq
	\langle T^{\mu\nu}(x_1)\hdots T^{\rho\lambda}(x_n)\rangle = 0\,.
\eeq
Our exposition is admittedly brief. We refer the reader to e.g., \cite{1991PhR...209..129B,Labastida:1997pb,Ivancevic:2008wf} for a thorough account of topological quantum field theories. Obviously, one may replace the metric and stress tensor in the derivation above with any other source and conjugate operator. For example, an external flavor field and its associated globally conserved current.

One means of generating the requirements of a Witten-type topological theory is to use superspace~\cite{Horne:1988yn}. Indeed, let us introduce a Grassmanian coordinate $\theta$ as one does in supersymmetric quantum mechanics, collecting ordinary fields into superfields which depend both on $x$ and $\theta$ and by arranging for $\delta_Q$ to act on superfields as a derivative in the $\theta$-direction. More explicitly, a superfield $\super{\Phi}$ may be expanded in the form
\begin{equation}
	\super{\Phi}(x,\theta) = \phi(x) + \theta \psi(x),
\end{equation}
whereby
\beq
\delta_Q \super{\Phi} = \frac{\partial}{\partial \theta} \super{\Phi} = \psi(x) \,.
\eeq
Note that $\psi$ is $Q$-exact; $\delta_Q \phi = \psi$, so that $\delta_Q \psi = 0$. 

Let us now work under the assumption that our cohomology is trivial, i.e., all $Q$-closed operators are exact. We then group all non-exact operators and their descendants into superfields. Our construction ensures that products of superfields are superfields, the bosonic derivative of a superfield is a superfield and superspace derivatives of superfields are also superfields. Thus, a super-Lagrangian which is local in the superfields will be $Q$-closed via the standard argument,
\beq
\label{E:SisQclosed}
	S = \int d^dx d\theta\, \super{L}\,, \qquad \delta_Q S = \int d^dx d\theta\, \frac{\partial \super{L}}{\partial \theta} = 0\,.
\eeq
For example, 
\begin{align}
\begin{split}
\label{E:SUSYexample}
	S &=  \int  d^dxd\theta \, \left(\frac{1}{2}\partial_{\mu} \super{\Phi}\partial^{\mu} \super{\Phi} + V(\super{\Phi}) \right)=\int d^dx \, \left(\partial_{\mu}\psi \partial^{\mu}\phi +\psi V'(\phi)\right)
	\\
	&=\int d^dx  \, \delta_Q \left( \frac{1}{2}(\partial_{\mu}\phi)^2 + V(\phi)\right)\,.
\end{split}
\end{align}
(The astute reader will notice that our example above is somewhat unorthodox given that the action is fermionic. This can be ameliorated by making the replacement $\partial_{\mu} \super{\Phi} \partial^{\mu} \super{\Phi} \to \super{\Phi} \partial_{\mu} \super{\Phi} \partial^{\mu} \super{\Phi}$ with $\phi$ fermionic.)

 Finally, in order for the theory to be topological we need that the stress tensor is $Q$-exact. After minimally coupling the theory to a metric, we can define a super stress tensor via
\begin{equation}\label{E:varT}
	\delta_g S = \frac{1}{2} \int d^dx d\theta \sqrt{-g} \,  \super{T}^{\mu\nu}\delta g_{\mu\nu}\,,
\end{equation}
and identify the top component of the super stress tensor with the physical one. By construction, the top component of $\super{T}^{\mu\nu}$ is $Q$-closed. In our earlier example \eqref{E:SUSYexample} we find
\begin{align}
\begin{split}
\label{E:superT}
	\super{T}_{\mu\nu} =& \left[ \partial_{\mu}\phi \partial_{\nu}\phi - \eta_{\mu\nu} \left( \frac{(\partial \phi )^2}{2}+ V(\phi) \right) \right]
		\\
		& \qquad + \theta  \bigg[ \partial_{\mu}\psi \partial_{\nu}\phi + \partial_{\mu} \phi \partial_{\nu}\psi  - \eta_{\mu\nu} \left( \partial_{\rho}\psi \partial^{\rho} \phi + \psi V'(\phi) \right) \bigg] \,,
\end{split}
\end{align}
which indeed has the expected properties.

\subsubsection{Source-deformed topological field theories}
\label{SS:SDTQFT}

When the sources of the Schwinger-Keldysh theory are not aligned, the theory will no longer be topological. To account for such a requirement we need to construct a ``source-deformed topological theory,'' i.e., a theory which ceases to be topological once certain sources are turned on. To construct such theories it is helpful to first take a step back and consider \eqref{E:SisQclosed}. With $\delta_Q$ acting as $\partial/\partial\theta$, the integral of a super-Lagrangian $\super{L}$ will be $Q$-closed as long as $\super{L}$ is a superfield, e.g., it does not explicitly depend on $\theta$. One uses the same line of reasoning to argue that the momentum operator generates a symmetry as long as the Lagrangian does not explicitly depend on the position.

Even when a Lagrangian depends explicitly on position, one can define a spurionic translation symmetry, by promoting the position-dependent couplings to spurions. For example, the Lagrangian of a massive scalar field $\phi$ with a position-dependent mass,
\beq
	L = \frac{1}{2} (\partial\phi)^2+\frac{m^2(x)}{2}\phi^2\,,
\eeq
is not translation-invariant since $m^2(x)$ is inert under the momentum operator. However, we can define a spurionic translation symmetry if we allow $m^2(x)$ to transform in the same way as $\phi(x)$ under translations. Defined this way, the Lagrangian is always invariant under the spurionic translation. It is invariant under physical translations if and only if the background field $m^2(x)$ is translation-invariant.

In the same way we can define a spurionic supersymmetry under which the action is always invariant by allowing couplings to transform as superfields. This is the route by which one proves almost all supersymmetric non-renormalization theorems~\cite{Seiberg:1993vc}. For example, consider the Grassman-odd functional 
\beq
\label{E:SUSYexample2}
S' = \int d^dx d\theta \left( \frac{1}{2}\partial_{\mu}\super{\Phi}\partial^{\mu}\super{\Phi} + a \super{\Phi} + \theta a_{\bar{g}} \super{\Phi} \right)=\int d^dx \left( \partial_{\mu}\psi\partial^{\mu}\phi + a \psi + a_{\bar{g}} \phi\right)\,,
\eeq
where $a$ is bosonic and $a_{\bar{g}}$ is Grassman-odd. It is equal to our example~\eqref{E:SUSYexample} when $a_{\bar{g}}=0$ and $V(\super{\Phi})=a \super{\Phi}$. The supersymmetry is broken by the explicit dependence on $\theta$. But $S'$ is invariant under a spurionic supersymmetry. Collecting  $a$ and $a_{\bar{g}}$ into a background superfield
\beq
	\super{A} = a + \theta a_{\bar{g}}\,,
\eeq
$S'$ becomes
\beq
	S' =\int d^dx d\theta \left( \frac{1}{2}\partial_{\mu}\super{\Phi}\partial^{\mu}\super{\Phi}  + \super{A} \super{\Phi} \right)\,.
\eeq
If we define a nilpotent operator $Q'$ such that 
\beq
	{\delta}_{Q'} \super{A} = \frac{\partial \super{A}}{\partial \theta}\,, \qquad {\delta}_{Q'} \super{\Phi} = \frac{\partial \super{\Phi}}{\partial\theta}\,,
\eeq
then clearly, $S'$ is $Q'$-closed.

The action $S'$ is also closed under $Q$ when the background superfield $\super{A}$ has no top component, or equivalently, when $\delta_{Q'} \super{A} = \frac{\partial \super{A}}{\partial\theta}=a_{\bar{g}}=0$. More generally, the action is $Q$-closed if all of the background superfields are ${\delta}_Q'$-invariant, meaning that all of their top components vanish. Put differently, a $Q$-closed action is one for whom all couplings are invariant under a superisometry generated by $\partial/\partial\theta$.\footnote{There is a similar statement for supersymmetric field theories coupled to a bosonic background of background supergravity. The supersymmetries which are preserved by the background, the superisometries, are those that leave the background invariant. See e.g.~\cite{Festuccia:2011ws}}

Observe that in~\eqref{E:SUSYexample2} the coupling $a$ acts as a source for the $Q$-exact operator $\psi$, while the supersymmetry-breaking coupling $a_{\bar{g}}$ acts as a source for the operator $\phi$. This is the prototype for a more general relation. Given any background superfield $\super{A} = a + \theta a_{\bar{g}}$, the variation of the action $S$ gives a conjugate super-operator $\super{O} = O+ \theta O_{\bar{g}}$ via
\beq
\delta S = \int d^dx d\theta  \, \super{O} \delta \super{A}= \int d^dx \left(  O_{\bar{g}}\delta a  +  O\delta a_{\bar{g}}\right)\,.
\eeq
Observe that the supersymmetry-preserving bottom component of $\super{A}$, $a$, couples to the $Q$-exact operator $O_{\bar{g}}$, while the supersymmetry-breaking top component, $a_{\bar{g}}$ couples to $O$.

This construction is exactly what we need. Grouping operators into superfields $\super{O}=O + \theta O_{\bar{g}}$, we can compute correlation functions of both $O$ and $O_{\bar{g}}$ by turning on a background superfield $\super{J}$. In this way the supersymmetry-breaking couplings are completely determined by the supersymmetry-preserving ones.

\subsubsection{Application to Schwinger-Keldysh partition functions}
\label{SS:applicationtoSK}

Let us apply the results of the previous subsections to the Schwinger-Keldysh effective action which must have a  topological sector  in the limit when the sources are aligned $A_1=A_2$. More precisely, correlation functions of the difference currents $J_a^i = J_1^i-J_2^i$ and stress tensor $T_a^{ij} = T_1^{ij}-T_2^{ij}$ must vanish whenever $B_{1\,\mu}(x)=B_{2\,\mu}(x)$ and $g_{1\,\mu\nu}(x) = g_{2\,\mu\nu}(x)$ (up to gauge transformations). %

We denote the supercharge associated with the topological Schwinger-Keldysh symmetry by $Q_{SK}$. We now require that the action $S_{eff}$ is $Q_{SK}$-exact whenever the sources are aligned. 
Recall that the dynamical fields $X$ and $C$ enter into the action only through the pulled back sources. Thus, we will collect the pulled back sources into multiplets such that the $a$-type pulled back sources become $Q_{SK}$-exact  sources are aligned.
Thus, to each $a$-type pulled back source $A_a$ we associate a ghost source $A_g$. Since $Q_{SK}$ annihilates $A_a = A_1 - A_2$, it follows that there is another ghost source $A_{\bar{g}}$ given by the action of $Q_{SK}$ on $A_r = (A_1+A_2)/2$. Together the $\{A_r,A_{\bar{g}}\}$ and $\{A_g,A_a\}$ make up the basic Schwinger-Keldysh supermultiplets which must obey the following algebra
\begin{align}
\begin{split}
\label{E:QSKrep}
	[Q_{SK}, \Rr A_r] &= \bGg A_{\bar{g}}\,, \hspace{1.48in} \{Q_{SK},A_{\bar{g}}\} = 0\,, \\
	\{Q_{SK}, \mathcal{G} A_g\} &= \Aa (A_1-A_2)= \Aa A_a\,, \hspace{.54in} [Q_{SK}, A_a] = 0\,.
\end{split}
\end{align}
Here $\{\Rr,\,\bGg,\,\Gg,\,\mathcal{A} \}$ are Grassmann-even operators that commute with $Q_{SK}$ which will be determined as we go along. In the remainder of this work we will consider bosonic $A_a$ and $A_r$ implying that the ghosts $A_g$ and $A_{\bar{g}}$ have Grassmann-parity $-1$. Apart from the ambiguity associated with $\{\Rr,\,\bGg,\,\Gg,\,\mathcal{A} \}$ we can also make the redefinition $A_g \to A_g + \mathcal{C}A_{\bar{g}}$. We will use this freedom in section \ref{S:KMS} when we treat the KMS symmetry.

Equation \eqref{E:QSKrep} specifies how $Q_{SK}$ should act on $A_r$ and $A_a$ and their respective ghost partners. But to properly define the action of $Q_{SK}$ we need to specify how it acts on the dynamical fields $X_1$, $X_2$, and $C_1$ and $C_2$, such that, e.g.,
\begin{equation}
\label{E:QSKBaexample}
	[Q_{SK},\, B_a] = [Q_{SK},\, X^{\mu}_1] \frac{\partial}{\partial X_1^{\mu}} B_1(X_1(\sigma)) - [Q_{SK},\, X^{\mu}_2] \frac{\partial}{\partial X_2^{\mu}} B_2(X_2(\sigma)) + \partial_i[Q_{SK},C_1]-\partial_i[Q_{SK},C_2]
\end{equation}
vanishes whenever the sources are aligned 
\begin{equation}
\label{E:aligned}
	B_{1\,\mu}(x) = B_{2\,\mu}(x) \equiv B_{\mu}(x)\,. 
\end{equation}	
If we work perturbatively around the aligned limit \eqref{E:aligned} then such a constraint may be imposed by setting
\begin{subequations}
\label{E:QSKonXC}
\begin{align}
\label{E:QSKonX}
	[Q_{SK},\,X_1^{\mu}] & = [Q_{SK},\,X_2^{\mu}]\,,  \\
\label{E:QSKonC}
	[Q_{SK},\,C_1] &= [Q_{SK},\,C_2] \,.
\end{align}
\end{subequations}
See \cite{Crossley:2015evo}. In this work we avoid a perturbative expansion around \eqref{E:aligned} by appealing to a probe limit, where the $X_1$ and $X_2$ fields and their ghost partners are ``frozen'' as in \eqref{E:probe}. Thus, \eqref{E:QSKonC} is sufficient to ensure that \eqref{E:QSKBaexample} is satisfied.

Geometrizing the action of $Q_{SK}$ as $\delta_{Q_{SK}} \to \frac{\partial}{\partial\theta}$, we combine the sources~$B_{1\,i}(\sigma)$ and $B_{2\,i}(\sigma)$ together with the dynamical fields $C_1$, $C_2$, $C_g$ and $C_{\bar{g}}$ into superfields $\super{B}_r$ and $\super{B}_a$ as

\beq
	\label{E:shortSKmultiplets}
	\super{B}_r = \Rr B_r +  \theta \bGg B_{\bar{g}}\,, \qquad
	\super{B}_a =  \Gg B_g -   \theta \Aa B_a\,,
\eeq
where
\begin{align}
\begin{split}
\label{E:Bs}
	B_{r\,i} &= \frac{1}{2}\left(B_{1\,i}+B_{2\,i}\right)
	\qquad
	B_{a\,i} = B_{1\,i}-B_{2\,i}
	\\
	B_{\bar{g}\,i} &= \partial_i C_{\bar{g}} 
	\hspace{.94in}
	B_{g\,i}  = \partial_i C_g\,,
\end{split}
\end{align}
and $B_{1\,i}$ and $B_{2\,i}$ were defined in \eqref{E:defBigi}.
If we add fictitious ghost sources $\beta_{\bar{g}\,i}$ and $\beta_{g\,i}$ (not to be confused with $\beta^i$) and promote $Q_{SK}$ to a spurionic symmetry then we may replace the last two equalities with
\begin{equation}
\label{E:spurionicB}
	B_{\bar{g}\,i} = \beta_{\bar{g}} + \partial_i C_{\bar{g}} \\
	\qquad
	B_{g\,i}  = \beta_g + \partial_i C_g\,.
\end{equation}
Observe that both the $a$-type fields $B_a$ and the ghosts $B_{\bar{g}}$ are $Q_{SK}$-exact and so comprise a topological sector. Moreover, $\super{B}_a$ and $\super{B}_r$ have opposite Grassman-parity.

In writing the various fields, we are working on the ``worldvolume'' $\WV$ as we discussed in section \ref{S:doubled symmetries}. The Schwinger-Keldysh effective action can then be written as the integral of a super-Lagrangian over $\theta$ and the bosonic worldvolume coordinates $\sigma^i$ as
\beq
	S_{eff} = \int d^d \sigma d\theta\,  \super{L}\,,
\eeq
where $\super{L}=L(\super{B}_r,\,\super{B}_a)$ is Grassman-odd. We pause here to make a brief but important remark.  The super-Lagrangian $\super{L}$ can depend on bosonic as well as on superspace derivatives  $\partial/\partial\theta$ of the superfields, both of which (anti)commute with $Q_{SK}$. Later, after treating the KMS symmetry, we will see that superspace derivatives will be modified. These modifications will be key to implementing dissipation when applying this formalism to thermal states.

We conclude this subsection with a few comments on ghosts. We have introduced the ghosts solely in order to ensure the existence of the topological limit. On account of the spin-statistics theorem, the ghosts are unphysical, and we expect there to be fundamental constraints on how they appear in the effective action. In the quantization of e.g., gauge theories and worldsheet string theory, one can prove no-ghost theorems for which ghost number conservation seems to be a necessity. While we do not yet endeavor to prove a no-ghost theorem for Schwinger-Keldysh effective theories,  we can define a sensible notion of ghost number following~\cite{weinberg1995quantum} by endowing $C_g$ with ghost number $1$ and $C_{\bar{g}}$ with ghost number $-1$. Assigning $\theta$ ghost number $1$, it follows that $\super{B}_r$ and $\super{B}_a$ have ghost number $0$ and $1$ respectively. We expect that we ought to impose a ghost number symmetry so that the total action will have ghost number zero, allowing terms like $\int d\theta \,\super{B}_a \super{B}_r$.

\subsection{The reality condition}
\label{SS:realitycond}

The next condition we wish to impose is the reality condition~\eqref{E:reality}, which we remind the reader, is given by
\begin{equation}
\label{E:Zreality}
	Z[A_1,A_2]^* =  Z[A_2^*,A_1^*]\,,
\end{equation}
and we have allowed for the possibility of complex sources conjugate to complex operators. In order to ensure \eqref{E:Zreality} we will impose a similar constraint on the effective action,
\beq
\label{E:realityonS}
	S_{eff}[\xi;\,A_1,\,A_2]^* = - S_{eff}[\xi';\,A_2^*,\,A_1^*]\,,
\eeq
where $\xi'$ is an appropriate transformation of the dynamical fields $\xi$. In the probe limit in which we are working in, the sources are given by $B_{1\mu}(x_1)$ and $B_{2\mu}(x_1)$, and the dynamical fields are the $C$'s and their ghost partners. 

The reality condition \eqref{E:Zreality} is a $\mathbb{Z}_2$ symmetry. In order to enforce it on the action it is convenient to construct an anti-linear operator $\Rho$ which acts on the sources via
\beq
	\Rho(B_1) = B_2^*
	\qquad
	\Rho(B_2) = B_1^*\,.
\eeq
We now extend the action of $\Rho$ onto the dynamical variables so that
\begin{equation}
\label{E:realitycondition2}
	S_{eff}[\super{B}_r,\,\super{B}_a]^* = -S_{eff}[\Rho(\super{B}_r),\,\Rho(\super{B}_a)]\,.
\end{equation}
Given that $S_{eff} = \int d^d\sigma d\theta \super{L}$, we find that
\beq
\label{E:realSKaction}
	S_{eff} = \int d^d\sigma d\theta \, L(\super{B}_a,\,\super{B}_r) = \int d^d\sigma \left(-\Aa B_a\left. \frac{\partial L}{\partial \super{B}_a}\right|_{\theta=0}  +\bGg B_{\bar{g}} \left. \frac{\partial L}{\partial \super{B}_r} \right|_{\theta=0}  \right)\,,
\eeq
and using \eqref{E:shortSKmultiplets} we find
\beq
\left. \super{B}_a\right|_{\theta=0} = \Gg B_g\,, \qquad \left. \super{B}_r\right|_{\theta=0} = \Rr B_r\,,
\eeq
where one has to be careful regarding signs when varying the fermionic Lagrangian with respect to fermionic operators. Since these signs will not play a role in our analysis, we omit them for brevity. In equation \eqref{E:realSKaction} and in the remainder of this section we will omit the dependence of $L$ on the sources for concicesness. Now, since  $\Rho$ is antilinear, i.e. 
\beq
	\Rho \left( i B \right) = -i \Rho (B)\,,
\eeq
we can use equation \eqref{E:realSKaction} to constrain $P$  such that 
 \begin{align}
 \begin{split}
 \label{E:R1}
	\Rho\left( B_r(\sigma) \right) &= ( B_r(\sigma) )^*\,, \hspace{.27in} 
	\Rho \left(  B_a(\sigma)  \right)  = -( B_a(\sigma) )^*\,, \\
	\Rho \left(  B_{\bar{g}}(\sigma) \right) & = -( B_{\bar{g}}(\sigma) )^*\,, \quad 
	\Rho \left(  B_g(\sigma) \right) =  ( B_g(\sigma) )^*\,.
 \end{split}
 \end{align}
 
In the remainder of this work we consider only real $C$'s and $B$'s. This implies that the ghosts $B_g$ and $B_{\bar{g}}$ are real Grassmannian fields. With \eqref{E:R1} an action of the form~\eqref{E:realSKaction} respects the reality condition~\eqref{E:realitycondition2} for real $L$ as long as $\Aa$, $\Gg$, $\bGg$ and $\Rr$ are real. We henceforth work with this convention. In their work, HLR \cite{Haehl:2015foa,Haehl:2015uoc,Haehl:2016pec}, christened $\Rho$ as a worldvolume CPT symmetry. However, we elect to follow CGL  \cite{Crossley:2015evo} and refer to it instead as a reality condition. At the end of the day this is just a semantic distinction. All groups demand that the effective action respects~\eqref{E:realitycondition2}.

In defining the action of $\Rho$ on the superfields we have the freedom of choosing how it acts on the superspace coordinate $\theta$.
For real $\Aa$ and $\bGg$ we may choose 
\begin{equation}
\label{E:realityoptionre}
	\Rho \left( \theta F \right) = -\theta \Rho (F)\,,
\end{equation}
for $F$ any external or dynamical field, so that $\Rho(\super{B}_a) = \super{B}_a$ and $\Rho(\super{B}_r) = \super{B}_r$.\footnote{Note that one may use an alternate convention where $\Aa$ and $\bGg$ are imaginary, in which case $L$ should be imaginary, and then a better choice for the action of $\Rho$ on superspace coordinates is $P \left( \theta F \right) = \theta P( F)$. } With these conventions we have
\beq
	\Rho( S_{eff}) = \int d^d\sigma \Rho(d\theta) \Rho( L(\super{B}_a,\super{B}_r)) = - \int d^d\sigma d\theta \,L^*(\Rho(\super{B}_a),\Rho(\super{B}_r)) = - \int d^d\sigma d\theta \, L^*(\super{B}_a,\super{B}_r)\,,
\eeq
This makes it clear that an action of the form~\eqref{E:realSKaction} will satisfy the reality condition~\eqref{E:realitycondition2} when $L$ is a real function of real supefields $\super{B}_a$ and $\super{B}_r$.

As discussed earlier, the super-Lagrangian $L$ may also contain superspace derivatives $\partial_{\theta}$. The most general action may be written as 
\beq
\label{E:realSKaction2}
S_{eff} = \int d^d\sigma d\theta\, L(\super{B}_a,\super{B}_r,\partial_i,i\partial_{\theta})\,.
\eeq
Under $\Rho$ we have 
\beq
\Rho(i\partial_{\theta}\super{B}_a) = i \partial_{\theta}\super{B}_a\,, \qquad \Rho(i\partial_{\theta}\super{B}_a) = i \partial_{\theta}\super{B}_a\,,
\eeq
so that
\beq
\Rho(S_{eff}) = -\int d^d\sigma d\theta\, L^*(\super{B}_a,\super{B}_r,\partial_i,i\partial_{\theta})\,.
\eeq
Equating the right-hand side with (minus)~\eqref{E:realSKaction2}, we see that the most general action~\eqref{E:realSKaction2}, with real $\super{B}'s$ respects the reality condition if and only if the super-Lagrangian $L$ is a real function.

We will see shortly that KMS symmetry effectively concatenates $\super{B}_a$ and $\super{B}_r$ into a long superfield $\super{B}$ and introduces a complex superspace derivative operator. We will revisit the effect of the reality condition on such long multiplets toward the end of Subsection \ref{Sec.KMS}.

\subsection{KMS symmetry}

Recall that the full KMS symmetry is a $\mathbb{Z}_2$ symmetry of the generating function \eqref{E:realZ2}. This $\mathbb{Z}_2$ symmetry leads to an additional topological sector which we referred to as a topological KMS symmetry. See \eqref{E:topologicalKMS}. In what follows we found it more convenient to construct the effective action by first implementing the topological KMS symmetry and then the full KMS symmetry. 

\subsubsection{Implementing the topological KMS symmetry}
\label{Sec.KMS}

Recall that the topological KMS sector arises whenever the $\wt{A}_a = (A_1 - e^{-i\delta_{\beta}}A_2)$ type sources vanish. To ensure the existence of an extra topological sector, we will use the same algorithm as the one discussed in section \ref{SS:applicationtoSK}: we assume the existence of a nilpotent linear operator which we  denote by $\overline{Q}_{KMS}$ and its action on superfields by $\delta_{\overline{Q}_{KMS}}$. We now  
strive to construct an action which will be $\overline{Q}_{KMS}$-closed when all $\tilde{a}$-type sources vanish $\wt{B}_a=0$.

In order to construct proper supermultiplets compatible with both  KMS and Schwinger-Keldysh symmetries, let us first deduce the action of $\overline{Q}_{KMS}$ on the basic fields $\{ B_r, B_{\bar{g}},B_g,B_a \}$. By definition, in a state characterized by $\beta$ and $\Lambda_{\beta}$, $\left(B_1 -  e^{-i\delta_{\beta}} {B}_2 \right) $ is $\overline{Q}_{KMS}$-exact. Since $\overline{Q}_{KMS}^2=0$, it follows that there exist two ghosts $\wt{B}_{\bar{g}}$ and $\wt{B}_g$ such that the action of $\overline{Q}_{KMS}$  is given by
\beq
\label{E:QKMS1}
	\left[\overline{Q}_{KMS},\,\left( \wt{\Rr} B_1 \right) \right] = \left[\overline{Q}_{KMS},\, e^{-i\delta_{\beta}} \wt{\Rr} {B}_2 \right] =   \wt{G} \wt{B}_g \,, \quad 
	\left\{\overline{Q}_{KMS},   \wt{\bGg}  \wt{B}_{\bar{g}}  \right\} =  -  \wt{\Aa} \wt{B}_{a} \,.
\eeq
Here, as in \eqref{E:QSKrep}, there is a great deal of freedom in choosing, e.g., the overall normalization of the ghost terms. 
We assume that the new supercharge $\overline{Q}_{KMS}$ is invariant under both flavor gauge transformations and coordinate reparameterizations. Then~\eqref{E:QKMS1} implies
\beq
	\left[\overline{Q}_{KMS},   \wt{\Rr} B_2  \right] = e^{i \delta_{\beta}}\wt{\Gg} \wt{B}_g \,.
\eeq

To determine the action of $\overline{Q}_{KMS}$ on the ghosts we make two important assumptions. The first is that the ghosts $\widetilde{B}_g$ and $\wt{B}_{\bar{g}}$ are linearly related to the ghosts $B_g$ and $\bar{B}_g$. We motivate this assumption by noting that the $\overline{Q}_{KMS}$ symmetry did not involve the introduction of new bosonic fields. Rather, it involved a linear combination of thermally shifted fields.

The second assumption is that the topological KMS and Schwinger-Keldysh symmetries remain distinct even in the high-temperature limit, $\delta_{\beta}\rightarrow 0$. Naively, we might expect that the opposite is true, on account of the fact that the $\overline{Q}_{KMS}$-exact operators $\wt{B}_a$ coincide with the $Q_{SK}$-exact operators $B_a$ at high temperature. Our motivation for taking $Q_{SK}$ and $\overline{Q}_{KMS}$ to be distinct is that, even when $\delta_{\beta}\to 0$, the full KMS symmetry~\eqref{E:KMSduality} remains non-trivial.

Let us start by implementing the expectation that $\widetilde{B}_g$ and $\wt{B}_{\bar{g}}$ are linear combinations of the $B_g$ and $B_{\bar{g}}$ ghosts: 
\beq
\label{E:theF}
\begin{pmatrix} \wt{B}_g \\ \wt{B}_{\bar{g}} \end{pmatrix} = \mathcal{F} \begin{pmatrix} B_g \\ B_{\bar{g}} \end{pmatrix}\,, \qquad \mathcal{F} = \begin{pmatrix} \mathcal{F}_g & \mathcal{F}_{\bar{g}} \\ \overline{\mathcal{F}}_{g} & \overline{\mathcal{F}}_{\bar{g}} \end{pmatrix}\,,
\eeq
with $\mathcal{F}$ an invertible matrix whose components are functions of $i\delta_{\beta}$.
We can always define a new barred ghost $\wt{B}_{\bar{g}}'$,
\begin{equation}
\label{E:tildegambiguity}
	\wt{B}_{\bar{g}} = \widetilde{B}_{\bar{g}}' + \mathcal{C}_1 \widetilde{B}_g\,,
\end{equation}
such that the relations \eqref{E:QKMS1} are still valid when replacing $\wt{B}_{\bar{g}}'$ with $\wt{B}_{\bar{g}}$. If $\mathcal{F}_g = 0$ we can use this freedom to set $\overline{\mathcal{F}}_{\bar{g}} = 0$. Otherwise, we use it to set $\overline{\mathcal{F}}_{{g}}=0$.
Thus, we have
\begin{equation}
\label{E:Fintermediate}
	\mathcal{F} = \begin{pmatrix} \mathcal{F}_g & \mathcal{F}_{\bar{g}} \\ 0 & \overline{\mathcal{F}}_{\bar{g}} \end{pmatrix}
	\quad
	\hbox{or}
	\quad
	\mathcal{F} = \begin{pmatrix} 0 & \mathcal{F}_{\bar{g}} \\ \overline{\mathcal{F}}_{{g}} & 0 \end{pmatrix}\,.
\end{equation}	

We now require that in the limit where $e^{i\delta_{\beta}}=1$ we get a non-trivial supercharge $\overline{Q}_{KMS}$. If we choose the second option in \eqref{E:Fintermediate} and set $\mathcal{F}_{\bar{g}}$ and $\mathcal{F}_g$ to constants then the relations \eqref{E:QKMS1} imply that $\overline{Q}_{KMS} \propto Q_{SK}$. So we choose the first possibility in \eqref{E:Fintermediate}. We can now use a redefinition of the untilde'd ghosts, ${B}_{{g}} = {B}_{{g}}' + \mathcal{C}_2 B_{\bar{g}}$ (similar to \eqref{E:tildegambiguity}) to set $\mathcal{F}_{\bar{g}}=0$ if it is not zero already. Thus, $\mathcal{F}$ is characterized by two functions which we choose to parameterize by $\mathcal{F}_g$ and $\det (\mathcal{F}) = \overline{\mathcal{F}}_{\bar{g}} \mathcal{F}_g$.

Given a pair of short superfields $\super{B}_a$ and $\super{B}_r$, equation \eqref{E:QKMS1} implies that $\overline{Q}_{KMS}$ mixes the components of $\super{B}_r$ and $\super{B}_a$ with each other. Thus, to realize the KMS topological symmetry we introduce an extra superspace coordinate $\btheta$ and concatenate $\super{B}_r$ and $\super{B}_a$ into a long superfield $\super{B}$,
\beq
\label{E:longmultiplet}
	\super{B} = \Rr B_r + \theta \bGg B_{\bar{g}} + \btheta \Gg B_g + \btheta \theta \Aa B_a  \,.
\eeq
Not any collection of operators of the form \eqref{E:longmultiplet} constitute a superfield. In order for $\super{B}$ to constitute a superfield we require that $Q_{SK}$ and $\overline{Q}_{KMS}$ act on it in a geometric way. Clearly
$
		\delta_{Q_{SK}}  \super{B} = \frac{\partial}{\partial \theta}  \super{B}\,.
$
One can check that in order for $\delta_{\overline{Q}_{KMS}}$ to act as a superdifferential operator on $\super{B}$ we must tune
\beq
\label{E:detF}
	 \frac{e^{-i\delta_{\beta}} \bGg \Gg \wt{\Aa} \wt{\Rr} }{\wt{\bGg} \wt{\Gg} \Aa \Rr \det(\mathcal{F})}  = 1\,,
\eeq
which gives us
\begin{align}
\begin{split}
\label{E:deltacomplicated}
	\delta_{Q_{SK}}  \super{B} &= \frac{\partial}{\partial \theta}  \super{B}\,, \\
	\delta_{\overline{Q}_{KMS}}  \super{B} &= \left( \frac{ 1}{2 }  \frac{\wt{\Gg} \Rr \mathcal{F}_g}{\Gg \wt{\Rr} } \left(1+ e^{i\delta_{\beta}}\right) \frac{\partial}{\partial \btheta} -   \frac{\wt{\Gg} \mathcal{F}_g \Aa }{\Gg \wt{\Rr}}  \left(1- e^{i\delta_{\beta}}\right) \theta \right)  \super{B}\,.
\end{split}
\end{align}

We can now choose 
\begin{equation}
\label{E:Fg}
	\frac{ 1}{2 }  \frac{\wt{\Gg} \Rr \mathcal{F}_g}{\Gg \wt{\Rr} } \left(1+ e^{i\delta_{\beta}}\right) =1\,,
\end{equation}
so that the derivative term in \eqref{E:deltacomplicated} takes a canonical form, and
\begin{equation}
\label{E:Avalue}
	2 \frac{\Aa}{\Rr} \frac{1-e^{i\delta_{\beta}}}{1+e^{i\delta_{\beta}}} = - i\delta_{\beta}\,,
\end{equation}
in order for $\delta_{\overline{Q}_{KMS}}$ to satisfy the Leibniz rule. In \eqref{E:Avalue} we have also chosen $\Aa/\Rr \to 1$ in the $\delta_{\beta} \to 0$ limit for later convenience. With these choices we find that the Schwinger-Keldysh and KMS variations are
\beq
\label{E:deltasimple}
	\delta_{Q_{SK}} \super{B} = \frac{\partial}{\partial \theta} \super{B}\,,
	\qquad
	\delta_{\overline{Q}_{KMS}} \super{B} = \left( \frac{\partial}{\partial \btheta} + i \delta_{\beta} \theta \right) \super{B}\,,
\eeq
for bosonic operators. Likewise,
\begin{equation}
\label{E:comsimple}
	\{Q_{SK},\,\overline{Q}_{KMS}\} =  i \delta_{\beta}  \,.
\end{equation}
It is interesting to note that this is the algebra of minimal supersymmetric quantum mechanics, with the thermal direction playing the role of time. The anticommutator \eqref{E:comsimple} has also been obtained by CGL \cite{Crossley:2015evo} and HLR \cite{Haehl:2015foa,Haehl:2015uoc}. In \cite{Crossley:2015evo} the authors seem to use $\{\delta_{Q_{SK}},\,\delta_{\overline{Q}_{KMS}}\} = 2 \tanh\left(i\delta_{\beta}/2\right)$. One way to obtain this result would be to use \eqref{E:detF}, \eqref{E:Fg} and $-\Aa=\Rr=1$. We refrained from doing so in order that the KMS supercharge satisfy the Leibniz rule. The authors of \cite{Haehl:2015foa,Haehl:2015uoc} have used $\{\delta_{Q_{SK}},\,\delta_{\overline{Q}_{KMS}}\} = 1 - e^{-i\delta_{\beta}}$, although they have two additional supercharges as we discuss in section~\ref{S:Discussion}.

Apart from the superfields given in \eqref{E:longmultiplet}, one can also generate  superfields from suitable superderivatives acting on~\eqref{E:longmultiplet}. Indeed, it is easy to check that
\beq
	\label{E:superDerivatives}
	D_{\theta} = \frac{\partial}{\partial \theta}  -  i \delta_{\beta} \btheta\,, \qquad D_{\btheta} = \frac{\partial}{\partial\btheta}\,,
\eeq
satisfy
\beq
\label{E:goodD}
	\{Q_{SK},D_{\theta}\} = \{Q_{SK},D_{\btheta}\} = \{\overline{Q}_{KMS},D_{\theta}\}=\{\overline{Q}_{KMS},D_{\btheta}\}=0\,,
\eeq
$D_{\theta}^2=D_{\btheta}^2=0$, and
\beq
\label{E:Dcom}
	\{D_{\theta},D_{\btheta}\} = - i \delta_{\beta}\,.
\eeq
Equation \eqref{E:goodD} ensures that $D_{\theta}\super{B}$ and $D_{\btheta}\super{B}$ are superfields.

It should be noted that $\super{B}$ was constructed by joining together the short superfields $\super{B}_a$ and $\super{B}_r$, which are the natural superfields associated with the topological Schwinger-Keldysh symmetry. One may, instead, consider superfields on which $\overline{Q}_{KMS}$ naturally acts, viz.,
\begin{equation}
\label{E:longtilde}
	\wt{\super{B}} = \wt{\Rr} \wt{B}_{r}   + \tilde{\theta} \wt{\bGg} \wt{B}_{\bar{g}}  + \tilde{\btheta} \wt{\Gg}  \wt{B}_{{g}} + \tilde{\btheta} \tilde{\theta} \wt{\Aa} \wt{B}_{a}\,,
\end{equation}
where the components of $\wt{\super{B}}$ have been defined in \eqref{E:defOaOr} and \eqref{E:theF}  and $\{ \wt{\Rr},\, \wt{\bGg},\, \wt{\Gg},\,\wt{\Aa} \}$ satisfy~\eqref{E:detF} and  \eqref{E:Fg}, and we use tilde'd superspace coordinates to distinguish them from the untilde'd ones. Indeed, one finds that 
\begin{align}
\label{E:tildeddelta}
	\delta_{Q_{SK}} \wt{\super{B}}  =  \left( \frac{\partial}{\partial \tilde{\theta}} +  i \tilde{\btheta}\delta_{\beta}   \right) \wt{\super{B}}\,,
	\qquad
	\delta_{\overline{Q}_{KMS}} \wt{\super{B}} = \frac{\partial}{\partial \tilde{\btheta}} \wt{\super{B}}\,,
\end{align}
where we have set
\begin{equation}
	2 \frac{\wt{\Aa}}{\wt{\Rr}} \frac{1-e^{i\delta_{\beta}}}{1+e^{i\delta_{\beta}}} = - i \delta_{\beta}\,.
\end{equation}
The natural superderivatives which act on the $\wt{\super{B}}'s$ are
\beq
\label{E:tildedDs}
	\wt{D}_{\tilde{\theta}} = \frac{\partial}{\partial \tilde{\theta}}\,, \qquad \wt{D}_{\tilde{\btheta}} = \frac{\partial}{\partial\tilde{\btheta}} - i \tilde{\theta} \delta_{\beta}\,.
\eeq
A priori, it would seem that we can choose whether to work with the $\super{B}$ superfields or the $\wt{\super{B}}$'s. In what follows we will use $\super{B}$. However, we forewarn the reader that in section \ref{SS:localKMS} we will see that the full KMS symmetry of the generating function forces us to use both types of superfields. 

Recall that (in the probe limit) it is $C_1$, $C_2$, $C_g$ and $C_{\bar{g}}$ which are dynamical and that these fields always appear in combination with the sources via \eqref{E:defBigi} and \eqref{E:Bs}. Thus, in order to ensure \eqref{E:deltasimple} we require that
\begin{equation}
	\delta_{Q_{SK}}\super{C} = \frac{\partial}{\partial \theta} \super{C},
	\qquad
	\delta_{\overline{Q}_{KMS}}\super{C} = \left(\frac{\partial}{\partial \btheta} + i\delta_{\beta}\theta\right)\super{C}\,.
\end{equation}
where
\be
\super{C}=\frac{1}{2}{\cal R}\left(C_1+C_2\right)+\bar{\theta}{\cal G}C_{g}+\theta{\bar{\cal G}}C_{\bar{g}}+\bar{\theta}\theta {\cal A}\left(C_1-C_2\right)
\ee
As was the case for the Schwinger-Keldysh symmetry, the KMS symmetry can be enhanced to a spurionic symmetry using \eqref{E:spurionicB}.

The most general Schwinger-Keldysh effective action is now given by the super-integral of a long superfield
\beq
\label{E:finalSuperActions}
S_{eff} = \int d^d\sigma d\theta d\btheta\, \super{L}\,.
\eeq
The super-Lagrangian $\super{L}$ may be constructed from superfields and their bosonic or superspace derivatives (see \eqref{E:superDerivatives}).

Let us now pause to revisit the reality condition for the effective action \eqref{E:realitycondition2},
\beq
  S_{eff}[\super{B}_r,\,\super{B}_a]^*=-S_{eff}[\Rho(\super{B}_r),\,\Rho(\super{B}_a)]\,,
\eeq
which in turn ensures \eqref{E:reality}. Recall that the transformation law for the antilinear operator $\Rho$ was given by \eqref{E:R1}. If $\super{B}$ is real then we maintain that for any component $B$ of $\super{B}$,
\begin{subequations}
\label{E:Preagain}
\beq
	P \left( \theta B \right) = -\theta P( B)\,,
	\qquad
	P \left( \btheta B \right) = \btheta P (B)\,,
\eeq
\begin{equation}
\label{E:Prerules}
	\Rho (\super{B}) = \super{B}\,,
	\qquad
	\Rho \left(D_{\theta} \super{B} \right) = -D_{\theta} \super{B}\,,
	\qquad
	\Rho \left(D_{\btheta} \super{B} \right) = D_{\btheta} \super{B}\,.
\end{equation}
\end{subequations}
Following the same logic we used at the end of Subsection~\ref{SS:realitycond}, we find that the most general action
\beq\label{E:seffder}
S_{eff} = \int d^d\sigma d\theta d\btheta\, L(\super{B},\partial,i D_{\theta},D_{\btheta};\beta,\Lambda_{\beta})\,,
\eeq
respects the reality condition~\eqref{E:realitycondition2}  if (and only if) $L$ is a real function of its arguments.

\subsubsection{The full KMS symmetry}
\label{SS:localKMS}

The topological KMS symmetry does not ensure that the full KMS symmetry \eqref{E:KMSv2} (or equivalently \eqref{E:KMSduality}) is satisfied. In what follows, we will ensure that the full KMS symmetry is satisfied by requiring that the effective action is invariant under an appropriate shift of its fields. In particular, following CGL \cite{Crossley:2015evo}, we demand that the effective action satisfy
\beq
\label{E:localKMS}
	S_{eff}[C,\,B_1,\,B_2;\beta,\Lambda_{\beta}] =  S_{eff}[\Theta^*C',\,\Theta^*B_1,\Theta^*e^{-i\delta_{\beta}}B_2;\Theta^*\beta,\Theta^*\Lambda_{\beta}]\,,
\eeq
where $C'$ is a suitable transformation of the dynamical fields $C$ that we will soon uncover. In terms of a Lagrangian (and temporarily ignoring the initial state data $\beta$ and $\Lambda_{\beta}$ to keep the presentation simple), we demand
\begin{align}
\nonumber
\int d^d\sigma \, L\left(C(\sigma),B_1(\sigma),B_2(\sigma),\partial\right) &= \int d^d\sigma\,L\left(\Theta^*C'(\sigma),\Theta^*B_1(\sigma),\Theta^*e^{-i\delta_{\beta}}B_2(\sigma),(\Theta^*)^2\partial\right)
\\
\label{E:localKMS2}
& = \int d^d\sigma \, \Theta^* L\left( C'(\sigma),B_1(\sigma),e^{-i\delta_{\beta}}B_2(\sigma),\Theta^*\partial\right)
\\
\nonumber
& = \int d^d\sigma\, L\left(\eta_{C} C'(\sigma),\eta_B B_1(\sigma),\eta_B e^{-i\delta_{\beta}}B_2(\sigma),\Theta^*\partial\right)\,,
\end{align}
where we have defined
\beq
	\Theta^* \frac{\partial}{\partial {\sigma} } = \frac{\partial} {\partial \vartheta {\sigma}}\,.
\eeq
In the last equality of \eqref{E:localKMS2} we have carried out a change of integration variables and $\eta_{C}$ is the eigenvalue of the dynamical field $C$ under \CPT and $\eta_B$ was defined in \eqref{E:defThetaA}.

The Lagrangian $L(C,\,B_1,\,B_2;\,\beta,\,\Lambda_{\beta})$ is, generically, not invariant under \eqref{E:localKMS2}. A prescription for making it invariant would be to modify the action by making the replacement
\beq
\label{E:KMSaction}
	L({\psi;\beta,\Lambda_{\beta}}) \to \frac{1}{2}\left(L({\psi;\beta,\Lambda_{\beta}}) +  L( \KMS \psi;\KMS\beta,\KMS\Lambda_{\beta} ) \right) \,,
\eeq
where we have collected the external and dynamical fields into $\psi$ and $\KMS$ is a linear $\mathbb{Z}_2$ transformation which depends on $\beta$ and $\Lambda_{\beta}$. In order for the total action
 to satisfy \eqref{E:localKMS2}, $\KMS$ acts on sources and derivatives as
\beq
\label{E:KactionS}
	\KMS B_1(\sigma) = \eta_B B_1(\sigma) \,,
	\qquad
	\KMS B_2(\sigma) = e^{-i \delta_{\beta}} \eta_B B_2(\sigma) \,,
	\qquad
	\KMS \frac{\partial}{\partial\sigma} = \Theta^* \frac{\partial}{\partial\sigma}\,,
\eeq
and on the thermodynamic data as
\beq
\label{E:Kactions}
	\KMS \beta^i = \eta_{\beta^i} \beta^i\,, 
	\qquad
	\KMS \Lambda_{\beta} =  -\Lambda_{\beta} \,,
\eeq
(recall that $\Lambda_{\beta}$ is odd under \CPT, and $\eta_{\beta^i}$ is the eigenvalue of $\beta^i$ under \CPT). 
Using \eqref{E:KactionS} and \eqref{E:Kactions} we find that $K \delta_{\beta} = - \delta_{\beta}$ implying, for example, that $\KMS^2 B_2(\sigma) = \KMS \left( \eta_Be^{-i \delta_{\beta} }B_2(\sigma) \right) = \eta_Be^{i\delta_{\beta}}K\left( B_2(\sigma)\right) =B_2(\sigma)$. Therefore, $K$ is indeed a $\mathbb{Z}_2$ transformation, squaring to one when acting on $(B_1,\,B_2;\beta^i,\Lambda_{\beta})$. In light of \eqref{E:KactionS} we also define the action of $K$ on the dynamical fields
\begin{equation}
\label{E:KactionO}
	\KMS C_1 = \eta_B C_1\,, 
	\qquad
	\KMS C_2 = \eta_B e^{-i \delta_{\beta}} C_2\,,
	\qquad
	\KMS \begin{pmatrix} C_g \\ C_{\bar{g}} \end{pmatrix}
	= 
	\eta_B \mathcal{S}(i\delta_{\beta})
	\begin{pmatrix} C_g \\ C_{\bar{g}}\end{pmatrix}\,,
\end{equation}
where $\mathcal{S}$ is a matrix. Since the action is bosonic, $K$ can square to either $1$ or $-1$ when acting on ghosts. In what follows, we use $K^2 = -1$ on ghosts. This condition implies that $\mathcal{S}$ satisfies $\mathcal{S}(-i\delta_{\beta})\mathcal{S}(i\delta_{\beta}) = -1$.

Observe that $K$ maps $r$-type and $a$-type operators to $\tilde{r}$-type and $\tilde{a}$-type operators respectively,
\beq
K C_r = \eta_B \wt{C}_r\,, \qquad K C_a = \eta_B \wt{C}_a\,,
\eeq
and vice versa. In order to ensure that K is manifestly consistent with the topological symmetries, its action on the ghosts and Grassmannian coordinates should be such that it maps superfields to superfields. Therefore, we define
\beq
\label{E:SUSYK}
	\KMS \super{C} = \eta_B \wt{\super{C}}\,,
\eeq
where the tilde'd multiplet $\wt{\super{C}}$ was defined in \eqref{E:longtilde}. The relation \eqref{E:SUSYK} may be ensured by requiring that
\beq
\label{E:Kthetarule}
	\KMS \left(\theta C\right) = -\tilde{\btheta}  \KMS C\,,
	\qquad
	\KMS \left(\btheta C\right) = \tilde{\theta} \KMS C\,,
\eeq
together with 
\beq
	\wt{\Rr}=\Rr\,,
	\qquad
	\wt{\Aa} = \Aa\,,
	\qquad
	\mathcal{S} = \begin{pmatrix} 0 & \frac{\bGg e^{-i\delta_{\beta}}}{\mathcal{F}_g \wt{\Gg} } \\ -\frac{\mathcal{F}_g \wt{\Gg}}{\bGg } & 0 \end{pmatrix}\,,
\eeq
which are compatible with \eqref{E:Avalue} and the requirement that $K$ is a $\mathbb{Z}_2$ transformation implies
\beq
	\left(\mathcal{F}_g \wt{\Gg} \right)^* = \mathcal{F}_g \wt{\Gg}  e^{i\delta_{\beta}}\,.
\eeq
Here we have used that $\bGg$ and $\Gg$ are real. Likewise, we find
\begin{align}
\begin{split}
\label{E:resultantD}
	\KMS \left(D_{\theta} \super{C}\right) =- \left( \frac{\partial }{\partial {\tilde{\btheta}} } - i \delta_{ \beta}  \tilde{\theta} \right) {\eta_B} \wt{\super{C}}\,
	= - \wt{D}_{\tilde{\btheta}} \left( \KMS {\super{C}} \right) \,,
	\qquad
	\KMS \left(D_{\btheta} \super{C}\right) = {\eta_B} \frac{\partial }{\partial {\tilde{\theta}}} \wt{\super{C}} =  \wt{D}_{\tilde{\theta}} \left( \KMS {\super{C}} \right) \,,
\end{split}
\end{align}
where the superderivatives on the right-hand side are the same ones we found in~\eqref{E:tildedDs}, appropriate when acting on tilde'd superfields.  

It is interesting to note that $K$ exchanges $Q_{SK}$ with $\overline{Q}_{KMS}$ and so is a sort of R-parity, in that, e.g.,
\beq
\label{E:Rsymmetry}
	K\left( \delta_{Q_{SK}}\super{C}\right) = -\delta_{\overline{Q}_{KMS}}\wt{\super{C}}\,, \qquad K\left(\delta_{\overline{Q}_{KMS}}\super{C}\right) = \delta_{Q_{SK}}\wt{\super{C}}\,.
\eeq
If we, once again, extend the supersymmetry to a spurionic one we may, for instance, extend \eqref{E:Rsymmetry} to the $\super{B}$ multiplets, 
\begin{equation}
	K\left( \delta_{Q_{SK}}\super{B}\right) = -\delta_{\overline{Q}_{KMS}}\wt{\super{B}}\,, \qquad K\left(\delta_{\overline{Q}_{KMS}}\super{B}\right) = \delta_{Q_{SK}}\wt{\super{B}}\,.
\end{equation}
In the remainder of this section we will work with spurionic supersymmetry.

Our final expression  for the effective action is now
\begin{align}
\begin{split}
\label{E:fullSeff}
	S_{eff}& = \frac{1}{2} \int d^d\sigma d\theta d\btheta L \left( \super{B},\,\partial,\,i D_{\theta},\,D_{\btheta} ;\, \beta,\,\Lambda_{\beta} \right) 
	\\
	& \qquad \qquad + \frac{1}{2} \int d^d\sigma d\tilde{\theta} d\tilde{\btheta} L \left( \eta_B \wt{\super{B}},\,\Theta^*\partial,\,-i \wt{D}_{\tilde{\btheta}},\,\wt{D}_{\tilde{\theta}} ; \, \eta_{\beta} \beta,\,-\Lambda_{\beta} \right)\,.
\end{split}
\end{align}
As discussed earlier, we have used tilde'd superspace coordinates in order to emphasize the distinction between tilde'd multiplets $\wt{\super{B}}$ and untilde'd ones $\super{B}$. 

It remains to check the compatibility of $\KMS$ with the reality condition. A short computation shows that 
\beq
\label{E:defRho}
	\Rho \left( \wt{\super{B}} \right) = e^{i\delta_{\beta}} \wt{ \super{B}}\,,
	\qquad
	\Rho \left( \wt{D}_{\tilde{\theta}} \wt{\super{B}} \right) = -e^{i \delta_{\beta}}\wt{D}_{\tilde{\theta}} \wt{\super{B}}\,,
	\qquad
	\Rho \left( \wt{D}_{\tilde{\btheta}}  \wt{\super{B}} \right)  = e^{i\delta_{\beta}} \wt{D}_{\tilde{\btheta}} \wt{\super{B}}\,,
\eeq
as long as
\beq
	\Rho\left(\tilde{\theta} B\right) = -\tilde{\theta} \Rho\left(B\right)\,,
	\qquad
	\Rho\left(\tilde{\btheta} B\right) = \tilde{\btheta} \Rho\left(B\right)\,.
\eeq
At first sight \eqref{E:defRho} seems at odds with \eqref{E:fullSeff}. Note however that a bosonic action will always have an even number of superderivatives. Put differently, $\wt{D}_{\tilde{\theta}}$ will always appear in conjunction with another $\wt{D}_{\tilde{\theta}}$ or with $\wt{D}_{\tilde{\btheta}}$ so that the reality condition is always satisfied.

Let us briefly dwell on ghost number. Providing $\btheta$ with ghost number $-1$ we find that $\super{B}$ has ghost number zero, $D_{\theta}$ has ghost number $-1$ and $D_{\btheta}$ has ghost number $1$. An action with ghost number zero allows for terms of the form $D_{\theta} \super{B}_1 D_{\btheta} \super{B}_2$ but not terms of the form $D_{\theta} \super{B}_1 D_{\theta} \super{B}_2$.

In the remainder of this work we will omit the tilde's on the superspace coordinates in the KMS partner action in order to tidy up our notation. We will also choose conventions where $\Aa \to 1$ (and therefore $\Rr \to 1$) in the $\delta_{\beta} \to 0$ limit and that $\Gg=\bGg=\wt{\Gg} = \wt{\bGg}=1$. Our convention for the ghosts enforces, via \eqref{E:detF} and~\eqref{E:Fg}, that
\beq
	\mathcal{F}_g=\frac{2}{1+e^{i\delta_{\beta}}}\,,
	\qquad
	\text{det}(\mathcal{F}) = e^{-i\delta_{\beta}}\,,
\eeq
and the tilde'd ghosts are
\beq
	\wt{B}_{\bar{g}} = \frac{1 + e^{-i \delta_{\beta}}}{2}B_{\bar{g}}\,,  
	\qquad
	\wt{B}_g = \frac{2}{1+e^{i\delta_{\beta}}}B_g \,.
\eeq
We provide a summary of our results in section \ref{S:action}.

\subsection{Ward identities}
\label{S:dof}

Recall that we have identified the mappings $x^{\mu}_1$, $x^{\mu}_2$, $c_1$ and $c_2$ with the dynamical fields $X^{\mu}_1$, $X^{\mu}_2$, $C_1$ and $C_2$ in order to ensure that the conservation equations emerge from equations of motion. In the current work we have focused on the probe limit where the only dynamical fields are $C_1$, $C_2$ and their ghost partners $C_g$ and $C_{\bar{g}}$. These were grouped into a superfield
\beq
\label{E:superC}
	\super{C} \equiv \Rr C_r + \theta C_{\bar{g}} + \btheta C_g + \btheta \theta \Aa C_a\,.
\eeq
The $\super{C}$ field is a scalar under worldvolume reparameterizations and transforms as a phase under worldvolume gauge transformations
\beq
\label{E:Ctransformation}
	\delta_{\Lambda}\super{C} = \Rr \Lambda\,.
\eeq

We now use $\super{C}$ and the trivial mapping $x^{\mu} = \delta_{i}^{\mu} \sigma^i$ to pull back target space sources to the worldvolume, 
\beq
\super{B}_i = \left[\frac{\Rr}{2}\left(B_{1\mu}(x(\sigma)) + B_{2\mu}(x(\sigma))\right) + \btheta \theta \Aa \left(B_{1\mu}(x(\sigma))-B_{2\mu}(x(\sigma)) \right)\right]\partial_i x^{\mu} + \partial_i \super{C}\,.
\eeq
Being constructed from ordinary functions of superfields, the super-pullbacks have the feature that they manifest the topological Schwinger-Keldysh and  KMS symmetries. In addition, this superfield is invariant under the target space transformations, 
\begin{align}
\begin{split}
	\label{E:WVU1}
		B_{1\mu} &\to B_{1\mu} + \partial_{\mu}\Lambda_1\,, \qquad
		B_{2\mu}  \to B_{2\mu} + \partial_{\mu}\Lambda_2\,, \\
		C_1 &\to C_1-\Lambda_1\,, \hspace{.63in}
		C_2  \to C_2 - \Lambda_2\,.
\end{split}
\end{align}

In the probe limit defined above it is particularly simple to show that the equations of motion for the dynamical variables are the Ward identities for the target space operators. Let us parameterize the variation of the effective action with respect to $\super{B}_i$, $\beta^j$, and $\Lambda_{\beta}$ as 
\begin{align}
\label{E:deltaSeff}
\delta S_{eff} &= \int d^d\sigma d\theta d\btheta \left\{  \left((\mathcal{R}\mathcal{A})^{-1} \super{J}^i \right) \delta \super{B}_i +\left( \mathcal{A}^{-1}\super{h}_i \right)\delta \beta^i + \left(\mathcal{A}^{-1}\super{m}\right) \delta \Lambda_{\beta}\right\}
\\
\nonumber
	& = \int d^d\sigma \left\{ J_r^i \delta B_{a\,i} + J_a^i \delta B_{r\,i} - \left( (\mathcal{R}\mathcal{A})^{-1}J_g^i\right) \delta B_{\bar{g}\,i} + \left( (\mathcal{R}\mathcal{A})^{-1}J_{\bar{g}}^i\right) \delta B_{g\,i} + h_{a\,i} \delta \beta^i + m_a \delta \Lambda_{\beta}\right\}\,,
\end{align}
where we have defined
\begin{equation}
	\super{J}^i = \Rr J_r^i + \theta J_{\bar{g}}^i + \btheta J_g^i + \btheta\theta \Aa J_a^i
	\qquad
	\hbox{and}
	\qquad
	\delta \super{B}_i= \Rr \delta B_{r\,i} + \theta \delta B_{\bar{g}\,i} + \btheta \delta B_{g\,i} + \btheta\theta \Aa \delta B_{a\,i}\,.
\end{equation}
We have added factors of $\Rr$ and $\Aa$ to the first line of \eqref{E:deltaSeff} in order that the bosonic terms in the second line take the standard form~\eqref{E:raBasis}. The first two entries in the variation indicate that $J_r^i$ and $J_a^i$ are indeed the worldvolume average and difference currents, and so we identify $\super{J}^i$ as the worldvolume super-$U(1)$ current. Without loss of generality we can promote $\beta^i$ and $\Lambda_{\beta}$ to be the bottom components of background superfields, conjugate to the superfields $\super{h}_i$ and $\super{m}$, although this is not necessary.

If we vary the target space sources $B_{1\,\mu}$ and $B_{2\,\nu}$ as well as the phase fields $C_1$ and $C_2$ but keep the ghost components of $C$ and the thermal data fixed, then \eqref{E:deltaSeff} becomes
\beq
\delta S_{eff} = \int d^d\sigma \left\{  J_r^{\mu}\delta B_{a\,\mu} +J_a^{\mu}\delta B_{r\,\mu} -E_r \delta C_a -E_a \delta C_r\right\} + (\text{boundary term})\,,
\eeq
with
\beq\label{E:pushforwards}
J_r^{\mu} = \delta_i^{\mu} J_r^i\,, \qquad J_a^{\mu} = \delta_i^{\mu} J_a^i\,, \qquad E_r =\partial_i J^i_a =  \partial_{\mu} J^{\mu}_a\,, \qquad E_a =\partial_iJ_r^i =  \partial_{\mu}J_r^{\mu}\,.
\eeq
We thereby identify the target space currents to be
\beq
J_1^{\mu} = \delta_i^{\mu} \left( J_r^i + \frac{J_a^i}{2}\right)\,, \qquad J_2^{\mu} = \delta_i^{\mu}\left( J_r^i - \frac{J_a^i}{2}\right)\,,
\eeq
and the equations of motion for $C_a$ and $C_r$, $E_r$ and $E_a$, are equivalent to the target space $U(1)$ Ward identities, which in the probe limit are simply
\beq
\partial_{\mu} J_1^{\mu} = 0\,, \qquad \partial_{\mu}J_2^{\mu} = 0\,.
\eeq

The perceptive reader might recall that  the action $S_{eff}$  need be invariant  under a worldvolume gauge symmetry defined in \eqref{E:Ctransformation}.
In our probe limit, the transformation law for the fields in the Lagrangian are
\beq
\delta_{\Lambda} \super{B}_i =\mathcal{R} \partial_i \Lambda\,.
\eeq
As we mentioned in~\eqref{E:Lbtransformation}, we let $\Lambda_{\beta}$ vary under worldvolume $U(1)$ transformations $\Lambda$ as 
\begin{equation}
\label{E:Lbtransform}
 \delta_{\Lambda}\Lambda_{\beta}  = -\beta^i \partial_i \Lambda\,.
\end{equation}
Inserting \eqref{E:Lbtransform} into~\eqref{E:deltaSeff} gives
\beq
\delta_{\Lambda} S_{eff} =- \int d^d\sigma \,\Lambda \left\{ \partial_i \left( J_a^i - \beta^i m_a\right)\right\}+(\text{boundary term})\,.
\eeq
We find that the  worldvolume gauge invariance implies that $\partial_iJ^i_a=\partial_i\left(\beta^i m_a\right)$ is identically satisfied. Once the equations of motion for the dynamical fields are imposed, $\partial_i J_a^i=0$, we see that necessarily $\partial_i (\beta^i m_a) =0$.

\section{Constructing the effective action}
\label{S:action}

Let us summarize our findings so far. The degrees of freedom of the effective hydrodynamic theory appear in the action through the superfields
\begin{align}
\begin{split}
	\super{B} &= \Rr B_r  + \theta B_{\bar{g}} + \btheta B_{{g}} + \btheta\theta \Aa B_{a}\,, \\
	\wt{\super{B}} &= \Rr \wt{B}_{r}  + \theta \wt{B}_{\bar{g}} + \btheta \wt{B}_{g} + \btheta\theta \Aa \wt{B}_{a}\,, \\
\end{split}
\end{align}
where we have defined the operators 
\begin{align}\label{E:def}
\begin{split}
	\wt{B}_r &=  \frac{1}{2} \left( 1+ e^{-i\delta_{\beta}} \right) B_r + \frac{1}{4}  \left( 1- e^{-i\delta_{\beta}} \right) B_a\,,
	\qquad
	\wt{B}_g = \frac{2}{1+e^{i\delta_{\beta}}} B_g\,,  \\
	\wt{B}_{a} &=  \frac{1}{2} \left(1 + e^{-i\delta_{\beta}} \right) B_a + \left(1-e^{-i\delta_{\beta}} \right) B_r\,, 
	\hspace{.44in}
	\wt{B}_{\bar{g}} =  \frac{e^{-i\delta_{\beta}} + 1}{2}  B_{\bar{g}}\,,
\end{split}
\end{align}
and the coefficient functions $\Aa$ and $\Rr$ must be real and satisfy
\begin{subequations}
\label{E:ARconventions}
\begin{align}
\begin{split}
	\frac{\Aa}{\Rr} &= \frac{1}{2} \coth\left(\frac{i \delta_{\beta}}{2} \right) i\delta_{\beta}\,, \\
	\end{split}
\end{align}
and
\beq
	\Aa \xrightarrow[\delta_{\beta}\to 0]{} 1\,.
\eeq
\end{subequations}
With these definitions the (spurionic) action of the Schwinger-Keldysh and KMS supercharges on the above superfields is given by~\eqref{E:deltasimple} and~\eqref{E:tildeddelta} to be
\begin{align}
\begin{split}
	\delta_{Q_{SK}} \super{B} &= \frac{\partial}{\partial \theta} \super{B}\,, 
	\hspace{1.01in}
	\delta_{\overline{Q}_{KMS}} \super{B} = \left(\frac{\partial}{\partial \btheta} + i \delta_{\beta} \theta \right) \super{B}\,, \\
	\delta_{Q_{SK}} \wt{\super{B}}& = \left(\frac{\partial}{\partial \theta} + i \delta_{\beta} \btheta \right) \wt{\super{B}}\,, 
	\qquad
	\delta_{\overline{Q}_{KMS}} \wt{\super{B}} =\frac{\partial}{\partial \btheta} \wt{\super{B} }	\,,
\end{split}
\end{align}
and the associated superderivatives are given by~\eqref{E:superDerivatives} and~\eqref{E:tildedDs}, 
\begin{align}
\begin{split}
	D_{\theta}  \super{B} &= \left(\frac{\partial}{\partial \theta} - i \delta_{\beta} \btheta \right) \super{B}\,, 
	\qquad
	D_{\btheta} \super{B} = \frac{\partial}{\partial \btheta} \super{B}\,, \\
	\wt{D}_{\theta} \wt{\super{B}} &=\frac{\partial}{\partial \theta} \wt{\super{B}}\,,  
	\hspace{1.02in}
	\wt{D}_{\btheta}  \wt{\super{B}} = \left(\frac{\partial}{\partial \btheta} - i \delta_{\beta} \theta \right) \wt{\super{B} }\,.
\end{split}
\end{align}

As discussed, we will work in the probe limit where the only sources are external $U(1)$ flavor gauge fields $B_{1\,\mu}$ and $B_{2\,\nu}$. The dynamical degrees of freedom are combined into a superfield
\begin{equation}\label{E:fieldsC}
	\super{C} = \Rr C_r + \theta  C_{\bar{g}} + \btheta C_{g} + \btheta \theta \Aa C_a\,.
\end{equation}
One now uses the identity map $X_1^{\mu}=X_2^{\mu} =  \delta^{\mu}_i\sigma^i$ to pull back the sources to the worldvolume and group them with $\super{C}$ to obtain the superfield~\eqref{E:probe}\begin{equation}\label{E:sourcesB}
	\super{B}_i = \Rr B_{r\,i} +\theta \partial_iC_{\bar{g}}+\bar{\theta}\partial_iC_g+ \bar{\theta}\theta\Aa B_{a\,i}\,, 
\end{equation}
where
\begin{equation}
B_{1\,i} = \delta_i^{\mu}B_{1\,\mu} + \partial_i C_1\,,\qquad B_{2\,i} = \delta_i^{\mu}B_{2\,\mu} + \partial_i C_2\,,
\end{equation}
and the average and difference fields and sources on the worldvolume are
\begin{equation}
 B_{i\,r} =\frac{1}{2} \left(B_{i\,1}+B_{i\,2}\right)\,,\quad B_{i\,a} = B_{i\,1}-B_{i\,2}\,,\quad C_{r}=\frac{1}{2}\left(C_1+C_2\right)\,,\quad C_a=C_1-C_2\,.
 \end{equation}
The superfield $\super{B}_i$ is invariant under target space $U(1)$ gauge transformations and it transforms as a connection under a worldvolume $U(1)$ gauge transformation,
\begin{equation}
\label{E:WVU1onB}
	\delta_{\Lambda}\super{B}_i ={\cal R} \partial_i \Lambda\,.
\end{equation}
The tilde'd version of $\super{B}_i$ is given by
\begin{equation}
	\wt{\super{B}}_i = \Rr \wt{B}_{r\,i} +\theta \partial_i\wt{C}_{\bar{g}}+\bar{\theta}\partial_i\wt{C}_g+ \bar{\theta}\theta\Aa \wt{B}_{a\,i} \,,
\end{equation}
as in \eqref{E:def}.

The most general worldvolume gauge invariant action depends on $\super{B}_i$ and $\wt{\super{B}}_j$, their derivatives, superderivatives, and the  external thermodynamic parameters $\beta^i$ and $\Lambda_{\beta}$. The parameter $\beta^i$ transforms as a gauge invariant vector and $\Lambda_{\beta}$ transforms as  
\begin{equation}
\label{E:WVU1onLambdab}
	\delta_{\Lambda}\Lambda_{\beta} = - \beta^i \partial_i \Lambda
\end{equation}
under the worldvolume $U(1)$ symmetry and as a scalar under coordinate transformations.

The Schwinger-Keldysh effective action is given by
\beq
     S_{eff} = \int d^d\sigma d\theta d\btheta \,\super{L}
\eeq
where
\beq
\label{E:fullaction}
	\super{L}=\frac{1}{2} L\left( \super{B},\, \partial,\,i D_{\theta},\,D_{\btheta} ;\, \beta ,\, \Lambda_{\beta} \right) + \frac{1}{2}L\left( \eta_{B} \wt{\super{B}},\,\Theta^* \partial,\, -i \wt{D}_{\btheta},\,\wt{D}_{\theta} ;\, \eta_{\beta} \beta,\,-\Lambda_{\beta} \right) 
\eeq
with $\Theta \partial_i = \partial / \partial \vartheta \sigma$ where $\vartheta \sigma$ is the \CPT \,transformation of $\sigma$, and $\eta_{B_i}$ and $\eta_{\beta^i}$ are the \CPT \,eigenvalues of $\super{B}_i$ and $\beta^i$ respectively. (In even dimensions we can set $\eta_{\beta^i} = 1$ and $\eta_{B_i}=-1$). The thermodynamic parameter $\Lambda_{\beta}$ has eigenvalue $-1$ under \CPT. We refer to the second term on the right as the KMS partner of $L$.

\subsection{The structure of the action}

 The appearance of superderivative terms in the action, or lack thereof, has interesting physical consequences. In essence, the superderivative terms control the number of $a$-type fields which appear in the bosonic action. We will see later, in section \ref{S:derivativeexpansion}, that the $a$-type fields correspond to what is often called stochastic noise in the context of dynamical equations coupled to a noise field. Such noise fields are often useful in providing for a description of time-dependent processes in dynamical critical phenomena. 

Let us begin our analysis by considering super-Lagrangians which contain no superderivatives. Omitting KMS partners, we have 
\begin{equation}
\label{E:Snod}
	\super{L} = {L}(\super{B},\partial;\beta,\Lambda_{\beta}) \,.
\end{equation}
A Lagrangian of the form \eqref{E:Snod} will contain only one power of the $a$-type fields after superspace integration. To see this, note that
\begin{equation}
	\int d\theta d\btheta \,\super{L}= \mathcal{A}B_{a\,i} \frac{\partial \super{L}}{\partial \super{B}_i}\Bigg|_{\theta=\bar{\theta}=0} +  \partial_iC_{\bar{g}} \, \partial_jC_{g} \frac{\partial^2 \super{L}}{\partial \super{B}_i \partial \super{B}_j} \Bigg|_{\theta=\bar{\theta}=0} \,.
\end{equation}
Put differently, the $r$-type current, $J_r^i$, that follows from \eqref{E:Snod} will contain no $a$-type fields. 

Let us now add superderivatives to the Lagrangian. The superderivatives of $\super{B}_i$ take the form
\begin{equation}
	D_{\theta} \super{B}_i =  \partial_iC_{\bar{g}} - \btheta  \frac{\Rr i\delta_{\beta}}{1-e^{-i\delta_{\beta}}} \wt{B}_{a\,i}-\btheta\theta  i \delta_{\beta} \partial_iC_{\bar{g}}\,,
	\qquad
	D_{\btheta} \super{B}_i =  \partial_iC_{g} + \theta \Aa B_{a\,i}\,,
\end{equation}
where we have used \eqref{E:def}. In order for the Lagrangian to be bosonic it must contain an even number of superspace derivatives. Given that $D_{\theta}^2 = D_{\btheta}^2=0$ the two derivative terms we can write are 
\begin{align}
\begin{split}
\label{E:simpleccterms}
	D_{\theta} \super{B}_i D_{\btheta} \super{B}_j & 
	= -\btheta\theta  \left(\frac{\Rr i \delta_{\beta}}{1-e^{-i\delta_{\beta}}} \wt{B}_{a\,i}  \right)\left(\Aa B_{a\,j} \right) + \left(\hbox{ghosts}\right)\\
	D_{\theta}\super{B}_i D_{\theta} \super{B}_j & = \left(\hbox{ghosts}\right) \\
	D_{\btheta}\super{B}_i D_{\btheta} \super{B}_j &  = \left(\hbox{ghosts}\right) \\
	D_{\theta}D_{\btheta} \super{B}_i & = \Aa B_{a\,i} - \btheta\theta \Aa i\delta_{\beta} B_{a\,i} + \left(\hbox{ghosts}\right)\\
	D_{\btheta}D_{\theta} \super{B}_i & = -  \frac{\Rr i \delta_{\beta}}{1-e^{-i\delta_{\beta}}} \wt{B}_{a\,i} + \left(\hbox{ghosts}\right)
\end{split}
\end{align}
where we have omitted ghost terms for brevity. Note that the last two terms in \eqref{E:simpleccterms} are related via $\{ D_{\theta},D_{\btheta}\} = - i \delta_{\beta}$, so we may omit one in place of the other. In what follows we will use the convention
\begin{equation}
	D \super{B}_i = D_{\theta} D_{\btheta} \super{B}_i\,.
\end{equation}

From \eqref{E:simpleccterms} and gauge invariance it follows that we may decompose the super-Lagrangian into scalar and tensor terms 
\begin{multline}
\label{E:superdecomposition}
	\super{L} =  \frac{1}{2}L(\super{B},\partial;\beta,\Lambda_{\beta}) +\frac{1}{2}\sum_{n=0} i^{n+1}\, L^{i j k_1 \ldots k_n }(\super{B},\partial;\beta,\Lambda_{\beta}) D_{\theta}\super{B}_{i}D_{\btheta} \super{B}_{j} D \super{B}_{k_1} \ldots D \super{B}_{k_n} \\
	+ \left(\hbox{ghost contributions}\right) + \left(\hbox{KMS partner}\right)  \,.
\end{multline}
A few comments are in order. We note that terms of the type $J^{k_1\ldots k_n} D\super{B}_{k_1} \ldots D \super{B}_{k_n}$ may be integrated by parts to yield the tensor terms appearing in \eqref{E:superdecomposition}, for this reason such terms have been omitted. We have seen that the scalar term in \eqref{E:superdecomposition} is of linear order in $a$. The tensor terms associated with $L^{ijk_1\ldots k_n}$ are of order $n+2$ in $a$-type fields. Note however that truncating \eqref{E:superdecomposition} to order $n$ does not reduce to an expansion in $a$-type fields due to the KMS partner action. 

The ghost contributions in \eqref{E:superdecomposition} involve terms which are undetermined from the bosonic part of the action. A subset of these must be set to zero when demanding that the action has ghost number zero. Terms which have ghost number zero and are not determined by the bosonic part of the action include, for instance, $D_{\theta} \super{B}_i D_{\btheta} \super{B}_j D_{\theta} \super{B}_k D_{\btheta} \super{B}_l$.

\subsection{The derivative expansion}
\label{S:derivativeexpansion}

Often, hydrodynamics is presented as a derivative expansion \cite{Baier:2007ix,Bhattacharyya:2008jc} where physical quantities are expanded in derivatives of the hydrodynamic variables. These are the temperature $T$, normalized velocity $u^{\mu}$, and chemical potential $\mu$.  In order to make contact with the hydrodynamic derivative expansion we must expand the effective action  \eqref{E:superdecomposition} in derivatives. 

Our conventions for counting derivatives are that the thermodynamic parameters $\beta^i$ and $\Lambda_{\beta}$ and the bottom component $B_{r\,i}$ are zeroth order in derivatives. Then $\delta_{\beta} =O(\partial)$, and 
\begin{align}
\begin{split}
	\frac{\Aa}{\Rr}  &= 1 + \frac{(i\delta_{\beta})^2}{12} + \mathcal{O}(\delta_{\beta}^4)
\end{split}
\end{align}
contains terms to all orders in the derivative expansion.
Thus, using \eqref{E:ARconventions}, the superfield
\begin{equation}
\label{E:Bexpandedind}
	\super{B}_i = B_{r\,i} + \theta \partial_i C_{\bar{g}} + \btheta \partial_i C_g + \btheta \theta B_{a\,i} + \mathcal{O}(\partial^2)\,,
\end{equation}
does too. Given that $D_{\theta}=\partial_{\theta} - i \btheta \delta_{\beta} $ we also need to associate a derivative counting to $D_{\theta}$ as well. We use conventions where $\theta$ and $\btheta$ are of order $\mathcal{O}(\partial^{-1/2})$ implying that $D_{\theta}$ and $D_{\btheta}$ are of order $\partial^{1/2}$ and $B_{a\,i}$ is first order in derivatives. Other conventions are also possible and will not modify the computation of measurable quantities. 

The KMS partner of $\super{B}_i$, $\wt{\super{B}}_i$, also contains an infinite power series in derivatives. This is expected. The fluctuation-dissipation theorem, which is enforced by the KMS symmetry, dictates an all-order in derivatives relation among correlation functions. Even if we defined the superfield $B_i$ to be $\super{B}_i = B_{r\,i} + \theta \partial_i C_{\bar{g}} + \btheta \partial_i C_g + \btheta \theta B_{a\,i}$ with no gradient corrections, ignoring the various problems that would arise with $\overline{Q}_{KMS}$ if we did so, its KMS conjugate $\wt{\super{B}}_i$ would have contained an infinite number of derivatives. With our current conventions, to first order in the derivative expansion, we have
\begin{align}
\begin{split}
\label{E:tildeBexpansion}
	\wt{\super{B}}_i =& \left( B_{r\,i} + \theta \partial_i C_{\bar{g}} + \btheta \partial_i C_g + \btheta \theta \left( B_{a\,i} + i \delta_{\beta} B_{r\,i} \right) \right) \\
	&- \frac{1}{2} i \delta_{\beta} \left( B_{r\,i} + \theta \partial_i C_{\bar{g}} + \btheta \partial_i C_g + \btheta \theta \left( B_{a\,i} + i \delta_{\beta} B_{r\,i}  \right) \right)
	+ \mathcal{O}(\partial^2)\,.
\end{split}
\end{align}

Consider a derivative expansion of the general action \eqref{E:superdecomposition}. Given such an expansion, one can compute the resulting current $\super{J}^i$ order by order in derivatives. In the absence of ghosts and $a$-type fields $J_r^i$, the bottom component of $\super{J}^i$, should satisfy the hydrodynamic constitutive relations appropriate at that order in the derivative expansion. In what follows we will show that this is indeed the case, up to first order in the derivative expansion. We defer a more extensive analysis to future work.

At zeroth order in derivatives, the only  gauge-invariant scalars we can construct out of $\beta^i$, $\Lambda_{\beta}$, and $\super{B}_i$ are
\beq
	\bbnu = \beta^i\super{B}_i + \Rr \Lambda_{\beta}\,,
	\qquad
	\hbox{and}
	\qquad
	T^{-2} = -\beta^i \beta_i\,,
\eeq
(see \eqref{E:WVU1onB} and \eqref{E:WVU1onLambdab} ). In what follows, we will use
\beq
	\bbnu = \Rr \,\nu + \theta \nu_{\bar{g}} + \btheta \nu_g + \btheta \theta \Aa (\beta^i B_{a\,i})\,,
\eeq
which implies that
\beq
	\nu =\beta^iB_{r\,i}+\Lambda_{\beta}\,, 
	\qquad
	\nu_{\bar{g}}= \beta^i\partial_iC_{\bar{g}}\,, \\ 
	\qquad
	\nu_{g}=\beta^i\partial_iC_g \,.
\eeq

At zeroth order in derivatives we can characterize the most general action \eqref{E:superdecomposition} by a single function $F(T,\nu)$:
\be
\label{eq:explicit:L0}
	S_{eff}=\frac{1}{2}\int d^d\sigma d\theta d\bt\,\bigg(F\left(T,\bbnu\right)+F\left(T, -\tilde{\bbnu}\right) \bigg) + \mathcal{O}(\partial) \,,
\ee
so that only the leading terms on the right-hand side of \eqref{E:Bexpandedind} and \eqref{E:tildeBexpansion} contribute at this order.

The supercurrent associated with this action is given by 
\begin{multline}
\label{E:supercurrent}
2 \left(J_r^i + \theta J_{\bar{g}}^i + \btheta J_g^i + \btheta \theta J_a^i \right)
		 = \left( \dot{F}\left(T,\nu\right) - \dot{F}\left(T,-\nu\right) \right) \beta^i 
		+ \left( \theta \nu_{\bar{g}} + \btheta \nu_g \right) \left(\ddot{F}(T,\nu)   + \ddot{F}(T,-\nu)   \right)  \beta^i  \\
		+  \btheta\theta \left( \beta^j B_{a\,j}  \left(\ddot{F}(T,\nu)   + \ddot{F}(T,-\nu)   \right)   + \nu_{\bar{g}} \nu_{{g}} \left(\dddot{F}(T,\nu)  - \dddot{F}(T,-\nu) \right)  \right) \beta^i
\end{multline}
where a dot denotes a derivative with respect to $\nu$. Thus,
\beq
\label{E:constitutiveO0V1}
	J_r^i = \frac{1}{2} \left( \dot{F}(T,\nu) - \dot{F}(T,-\nu) \right) \beta^i\,.
\eeq
The expected constitutive relations for a charge current, at zeroth order in derivatives, are 
\beq
\label{E:ConstitutiveO0V2}
	J_r^i = \rho  u^i\,,
\eeq
where $\rho$ is the charge density which is related to the pressure $P$ via the usual thermodynamic relation $\rho = \partial P / \partial \mu$. Recall that the pressure must be an even function of the chemical potential $\mu$ in order to retain CPT invariance of the theory \cite{Banerjee:2012iz}. As a result, $\rho$ must be an odd function of the chemical potential. Indeed, in comparing \eqref{E:constitutiveO0V1} with \eqref{E:ConstitutiveO0V2} we find that we may identify the velocity $u^i$, temperature $T$, and chemical potential $\mu$ as
\beq
T = \frac{1}{\sqrt{-\beta^2}} \,, \qquad u^i = \frac{\beta^i}{\sqrt{-\beta^2}}\,, \qquad \mu = T\nu\,,
\eeq
and, more importantly, the pressure $P$ with the the even part of $F$,
\beq
	P(T,\nu) = \frac{1}{2} \left( F(T,\nu) + F(T,-\nu) \right)\,.
\eeq
A similar expression arises for the charge density $\rho$. It is gratifying that the constitutive relations we obtain in \eqref{E:constitutiveO0V1} naturally respect CPT.

Let us now consider terms which are first order in derivatives. At that order, the effective action has the form
\beq
	S_{eff} = \frac{1}{2} \int d^d\sigma d\theta d\btheta \left( {L}_{0} + i\, L^{ij} D_{\theta} \super{B}_i D_{\btheta} \super{B}_j + \left(\hbox{KMS partners}\right)+ \left( \hbox{ghost terms} \right) \right)+ \mathcal{O}(\partial^2)\,,
\eeq
where the scalar contribution, $L_0$, has terms with at most one derivative and the tensor contribution $L^{ij}$ is zeroth order in derivatives. We have already seen that $L_0 = F + O(\partial)$. Corrections to the scalar action will come from scalar terms with one derivative. The possible gauge invariant scalars  with one derivative at our disposal are: 
\beq
\label{E:firstorder}
	\hbox{One derivative scalars:}
	\qquad
	\partial_i \beta^i\,,
	\qquad
	\beta^i \partial_i T\,,
	\qquad
	\beta^i \partial_i \bbnu \,.
\eeq
Since we are working in the probe limit, we have $\partial_i \beta^i = 0$ and $\beta^i \partial_i  T = 0$. Thus, the scalar part of the action can be written in the form:
\beq
\label{E:scalarL}
	{L}_{0} =  P(T,\bbnu) + p(T,\bbnu) \beta^i \partial_i  \bbnu \,,
\eeq
where we have, with some foresight, identified the pressure term from our analysis of the zeroth order derivative expansion. Note that the second term on the right-hand side of \eqref{E:scalarL} can be written as a total derivative and therefore will not contribute to the equations of motion or currents. 

As for $L^{ij}$, there are two tensor structures which we can write down at zeroth order in derivatives:\footnote{In $2+1$ dimensions there is another tensor structure available, $\epsilon^{ijk}\beta_k$. Including this structure and computing the ensuing response, one finds that this term leads to the anomalous Hall conductivity of~\cite{Jensen:2011xb}.}
\beq
\label{E:zeroorder}
	\hbox{Zero derivative tensors:} 
	\qquad
	P^{ij} \equiv \eta^{ij} +\frac{\beta^i \beta^j}{(-\beta^2)}\,,
	\qquad
	\beta^i \beta^j\,.
\eeq
The most general $L^{ij}$ we can write down is then given by
\beq
		{L}^{ij} = - \kappa(T,\bbnu)P^{ij} - s(T,\bbnu) \beta^i \beta^j  \,.
\eeq
Carrying out the superspace integration we find that, for the scalar Lagrangian,
\begin{align}
	\nonumber
	\int d\theta d\btheta \left( L_0 + \left(\hbox{KMS partner}\right)  \right) =&   B_{a\,i} \beta^i\left( \dot{P}(\nu) - \dot{P}(-\nu) \right) \\
	\label{E:scalarO1}
	&+ \frac{1}{2} i \delta_{\beta} \left( 2 P(-\nu) - i \delta_{\beta} P(-\nu)  + \beta^i B_{a\,i}  \dot{P}(-\nu)  \right) \\
\nonumber
	&+ \delta_{\beta}\left( \beta^i B_{a\,i} \left(p(\nu) + p(-\nu) \right) - i\delta_{\beta} \mathfrak{p}(-\nu) \right)
	+\left(\hbox{ghosts}\right)  \,,
\end{align}
where $\mathfrak{p}' = p$ and $\dot{P}(\nu) = dP/d\nu$. Note that the last two lines are a total derivative and do not contribute to the equations of motion or currents.

For the tensor Lagrangian we find
\begin{align}
	\nonumber
	\int d\theta d\btheta \left( i\, L^{ij} D_{\theta} \super{B}_i D_{\btheta} \super{B}_j  + \left(\hbox{KMS partner}\right)  \right) 
	=& i \left( \kappa(\nu) + \kappa(-\nu) \right)  P^{ij}  (B_{a\,i} + i \delta_{\beta} B_{r\,i}) B_{a\,j} 
	\\
	\nonumber
	&+i \left( s(\nu) + s(-\nu) \right) (\beta^i B_{a\,i} + i \delta_{\beta} \nu )\beta^j  B_{a\,j}\,,
	\\
	\label{E:tensorO1}
	& \qquad + (\text{ghosts})\,.
\end{align}
As should be clear from \eqref{E:scalarO1} and \eqref{E:tensorO1}, only the symmetric part of $P$, $\kappa$ and $s$ under $\super{\nu}\to - \super{\nu}$ will contribute to the constitutive relations. Therefore, without loss of generality we will set
\beq
	\label{E:parity}
	P(T,\nu) = P(T,-\nu)\,, 
	\qquad 
	\kappa(T,\nu) = \kappa(T,-\nu)\,,
	\qquad
	s(T,\nu) = s(T,-\nu)\,.
\eeq
We find (neglecting total derivatives and ghosts)
\beq
\label{E:finalAction}
	\int d\theta  d\btheta \super{L} = \left( \dot{P}  - s  \delta_{\beta} \nu \right) \beta^j B_{a\,j} - \kappa P^{ij}  \left( \partial_i \nu -G_{r\,ik}\beta^k \right)  B_{a\,j} + i \left( \kappa  P^{ij} B_{a\,i}B_{a\,j}  + s  \left(\beta^i B_{a\,i} \right)^2 \right)\,,
\eeq
where we have used
\beq
	\delta_{\beta} B_{r\,i} = \pounds_{\beta} B_{r\,i} + \partial_i \Lambda_{\beta} = -G_{r\,ij}\beta^j + \partial_i (\beta^j B_{r\,j} + \Lambda_{\beta}) = \partial_i \nu - G_{r\,ij} \beta^j\,,
\eeq
c.f., \eqref{E:deltabeta}.

Computing the $r$-type current in the absence of $a$-type sources and setting $a$-type fields and ghosts to vanish we find that 
\beq
\label{E:constitutive1}
	J_{r}^i = (\dot{P}  -s \beta^j \partial_j \nu )\beta^i - \kappa  P^{ij}  \left( \partial_j \nu -  G_{jk}\beta^k \right)\,.
\eeq
After a field redefinition of $\nu$ of the form $\nu \to \nu - \frac{s}{\ddot{P}} \beta^i \partial_i \nu$ we find that \eqref{E:constitutive1} describes the constitutive relations for a charged particle with conductivity $\sigma = \kappa/T$ in the Landau frame,
\beq
\label{E:constitutive2}
	J_{r}^i = \rho u^i + \sigma  P^{ij}  \left( E_j-T\partial_j \frac{\mu}{T}\right)\,,
\eeq
where $E_i = G_{ij}u^j$ is the average electric field. Note that due to \eqref{E:parity}, the conductivity $\sigma$ is even under $\mu\to - \mu$, as expected by CPT. Equation \eqref{E:constitutive2} describes a worldvolume current. To obtain the target space $r$-type current we can use the pushforwards described in \eqref{E:pushforwards} to obtain
\beq
\label{E:constitutive3}
	J_{r}^\mu = \rho u^\mu + \sigma  P^{\mu\nu}  \left( E_{\nu}-T\partial_{\nu} \frac{\mu}{T}\right)\,.
\eeq

The effective action~\eqref{E:finalAction} has nonzero imaginary part,
\beq
	\text{Im} \int d\theta d\btheta\,\super{L} =  \left( \kappa P^{ij} + s \beta^i \beta^j\right) B_{a\,i}B_{a\,j} + \mathcal{O}(\partial^3)\,,
\eeq
which may also be understood as the leading order contribution in a small$-a$ expansion also expanded in derivatives. In general, we require that $\text{Im}(S_{eff})$ is bounded below in order for the functional integral to converge. In and of itself, this does not lead to a constraint on $\kappa$ and $s$. 
Fortunately, the authors of~\cite{Liu} have recently proven that $\text{Im}(S_{eff}) \geq 0$ in a setup very similar to ours.
This implies
\beq
	\sigma \geq 0\,,
	\qquad
	\hbox{and}
	\qquad
	s \geq 0\,,
\eeq
which recovers the textbook result $\sigma =\kappa/T\geq 0$.

As discussed in \cite{Haehl:2015uoc,Crossley:2015evo}, the $a$-type fields in the effective action are associated with noise in stochastic hydrodynamics \cite{Kovtun:2014hpa,Harder:2015nxa}. Following HLR~\cite{Haehl:2015uoc}, we introduce a noise field $N_i$ via a Hubbard-Stratanovich transformation. Under the functional integral we have
\begin{align}
\begin{split}
	S_{eff} = \int d^d\sigma 
		&\left(  \left( \dot{P}  - s  \delta_{\beta} \nu \right) \beta^j  - \kappa P^{ij}  \left( \partial_i \nu -G_{r\,ik}\beta^k \right)  - 2 \left(s \beta^i \beta^j  +\kappa P^{ij}   \right) N_i \right)  B_{a\,j} 
		\\
		&\qquad \qquad+ i \left( \kappa  P^{ij} + s  \beta^i \beta^j \right) N_i N_j \,,
\end{split}
\end{align}
so that the constitutive relations, (in a non Landau frame) are given by
\begin{equation}
	J_{r}^i = \left(   \frac{\dot{P}}{T}  - \frac{s}{T^2} u^j \partial_j \nu \right) u^i + \sigma  P^{ij}\left( E_j-T\partial_j \frac{\mu}{T}  \right) - 2 \left(\frac{s}{T^2} u^i u^j + \sigma T P^{ij} \right) N_j\,,
\end{equation}
where $N_j$ is a random noise drawn from a Gaussian sample whose transverse components have inverse width proportional to the conductivity and longitudinal components proportional to $s$. As emphasized by CGL~\cite{Crossley:2015evo}, this rewriting in terms of stochastic hydrodynamics is only  valid when higher order $a$-type fields are neglected.
If we were to continue beyond quadratic order in the small$-a$ expansion, it would no longer be possible to account for the $a$-type fields with noise via a Hubbard-Stratonovich transformation. In general one needs the full Schwinger-Keldysh effective action to properly treat thermal fluctuations, through the $a$-type fields.

\subsection{The fluctuation-dissipation relation}

Instead of carrying out a derivative expansion to characterize the constitutive relations it is also possible to carry out an expansion of the action in fields. Such an expansion should give us a handle on exact relations among thermal correlation functions, such as the fluctuation-dissipation theorem.

Consider a quadratic action of the form 
\begin{equation}
	S_{eff} = \frac{1}{2} \int d^d\sigma d\theta d\btheta \left( \super{B}_i F^{ij}(\partial,\,\beta) \super{B}_j + i D_{\theta}\super{B}_i \sigma^{ij}( \partial,\,\beta) D_{\btheta} \super{B}_j +\left( \hbox{KMS partners} \right) \right)+ \mathcal{O}(\super{B}^3)\,,
\end{equation}
in an even number of spacetime dimensions, with real $F^{ij}$ and $\sigma^{ij}$. We assume that $F^{ij}$ is such that the action is gauge invariant. In order to simplify the ensuing computation let us  perform the Fourier transform on the fields
\be
	O(\sigma)=\int \frac{d^d k}{(2\pi)^d}e^{ik\cdot\sigma}O(k)=\int \frac{d\omega d^{d-1} \vec{k}}{(2\pi)^d}e^{-i\omega\tau+i \vec{k}\cdot \vec{\sigma}}O(\omega,\vec{k})\,.
\ee
We further simplify our analysis by going to the static gauge where $\beta^i\partial_i = b \partial_{\tau}$ and $\Lambda_{\beta}=0$, 
to obtain
\begin{align}
\begin{split}
\label{E:Fourierquadratic}
	S_{eff} &= \frac{1}{2} \int \frac{d^dk}{(2\pi)^d} d\theta d\bar{\theta}\bigg( \super{B}_i(k) F^{ij}(-ik,\beta) \super{B}_j(-k) + i D_{\theta}\super{B}_i(k) \sigma^{ij}(-ik,\,\beta) D_{\btheta} \super{B}_j(-k) \\
	& \qquad \qquad\qquad \qquad\quad + \wt{\super{B}}_i(k) F^{ij}(ik,\beta) \wt{\super{B}}_j(-k) -i \wt{D}_{\btheta}\wt{\super{B}}_i(k) \sigma^{ij}(ik,\,\beta) \wt{D}_{\theta} \wt{\super{B}}_j(-k) \bigg)  \,.
\end{split}
\end{align}

Expanding the action \eqref{E:Fourierquadratic} in terms of $a$ and $r$-type sources and omitting the ghost terms we find
\begin{multline}
\label{E:topcquadratic}
	S_{eff} = \int \frac{d\omega d^{d-1}{k}}{(2\pi)^d}  
				 B_{r\,i}\left(\omega,\,{k}\right) \Aa(b \omega) \Rr(b\omega) \\
		\times \left( \Real \left(F^{ij}(i\omega,\,ik)+ F^{ji}(i \omega,\,ik) \right) - \frac{i b}{2} \omega \left(\sigma^{ij}(-i\omega,\,-i k) + \sigma^{ji}(-i\omega,\,-i k) \right) \right) B_{a\,j}\left(-\omega,\,-{k}\right) \\
	- i B_{a\,i}\left(\omega,\,{k}\right) \Aa(b\omega)^2 \Real \left(\sigma^{ij}(i\omega,\,ik \right)) B_{a\,j}\left(-\omega,\,- {k}\right) \,,
\end{multline}
with $F^{ij} = F^{ij}(i\omega,\,ik)$ and $\sigma^{ij} = \sigma^{ij}(i\omega,ik)$.
In obtaining \eqref{E:topcquadratic} we have used that $\Aa$ and $\Rr$ are real and hence symmetric under a sign flip of their argument. The resulting (classical worldvolume) momentum space correlators are given by (see \eqref{E:rasToar})
\begin{align}
	\nonumber
	G_{sym}^{ij} & =  - i \Aa(b\omega)^2 \Real \left(\sigma^{ij}(i\omega,\,ik) + \sigma^{ji}(i\omega,\,ik)  \right) \,,
	\\
	\nonumber
	G_{adv}^{ij} & = i \Aa(b\omega) \Rr(b\omega) \left(\Real \left(F^{ij}(i\omega,\,ik)+F^{ji}(i\omega,\,ik) \right) + \frac{i b \omega}{2} \left(\sigma^{ij}(i\omega,\,ik) + \sigma^{ji}(i\omega,\,ik) \right) \right)\,,
	\\
	\nonumber
	G_{ret}^{ij} & = i\Aa(b\omega) \Rr(b\omega) \left(\Real \left(F^{ij}(i\omega,\,ik)+F^{ji}(i\omega,\,ik)\right) -\frac{ i b \omega }{2} \left(\sigma^{ji}(-i\omega,\,-ik) + \sigma^{ji}(-i\omega,\,-ik) \right) \right) \,,
	\\
	\label{E:Greens}
	G_{aa}^{ij} & = 0 \,.	
\end{align}

It is easy to see that the $G^{ij}$'s in~\eqref{E:Greens} satisfy
\begin{align}
\begin{split}
\label{E:FDT}
	i G_{sym}^{ij} &= \frac{1}{2} \coth\left(\frac{b\omega}{2} \right) \left(G^{ij}_{ret}-G^{ij}_{adv}\right)\,.
\end{split}
\end{align}
The result \eqref{E:FDT} is reminiscent of the fluctuation-dissipation theorem, but is somewhat misleading in that we have not yet solved the equations of motion for the $\super{C}$ fields. In an upcoming manuscript we will demonstrate that the relations \eqref{E:FDT} still hold after carrying out the Gaussian path integral and translating \eqref{E:Greens} to target space Green's functions.

\section{Discussion and outlook}
\label{S:Discussion}

In this work we have used the microscopic symmetries of Schwinger-Keldysh partition functions to determine the constraints on effective field theories for thermal states. We then used those constraints to obtain effective field theories for dissipative hydrodynamics. Our work is largely inspired by that of Haehl, Loganayagam, and Rangamani (HLR)~\cite{Haehl:2015foa,Haehl:2015uoc} and is also based on Crossley, Glorioso, and Liu (CGL)~\cite{Crossley:2015evo} although it differs from them in several respects.

\subsection{Comparison with previous work}

While our expression for the effective action is most similar to that of HLR \cite{Haehl:2015foa,Haehl:2015uoc} the details and interpretation differ significantly. For example, the authors of \cite{Haehl:2015foa,Haehl:2015uoc} consider actions which are invariant under four supercharges, as opposed to our two. Further, in addition to the fields we have described they advocate for a dynamical $U(1)_T$ field. HLR \cite{Haehl:2015foa,Haehl:2015uoc} also do not appear to use the KMS partner terms in the action, which we implemented in order for the fluctuation-dissipation relations to be satisfied. Let us address some of these points in detail.

As discussed in section \ref{S:SK}, the Schwinger-Keldysh partition function has a topological limit. We enforced this property by positing a single BRST-like supercharge $Q_{SK}$. The authors of~\cite{Haehl:2015foa,Haehl:2015uoc} have postulated that the topological limit is imposed through two such supercharges, $Q_{SK}$ and $\overline{Q}_{SK}$. We are not aware of a proof that there is always a second supercharge $\overline{Q}_{SK}$, nor of a counterexample in which it is forbidden.  The extra generator $\overline{Q}_{SK}$ is somewhat reminiscent of the generator of an anti-BRST symmetry which emerges in the BRST quantization of gauge theories \cite{Curci:1976yb,Baulieu:1981sb,Ojima:1980da}. Along with the topological Schwinger-Keldysh symmetry, we also imposed a topological KMS symmetry generated by $\overline{Q}_{KMS}$. This is at odds with the pair of KMS supercharges, $Q_{KMS}$ and $\overline{Q}_{KMS}$ posited by the authors of~\cite{Haehl:2015foa,Haehl:2015uoc}. 
In our work we account for $Q_{SK}$ and $\overline{Q}_{KMS}$ by introducing a superspace spanned by two Grassmannian coordinates. The authors of~\cite{Haehl:2015foa,Haehl:2015uoc} also introduce such a two parameter superspace. 

Another prominent feature of~\cite{Haehl:2015foa,Haehl:2015uoc} is the existence of a dynamical gauge field $A_I$ with $I$ running over worldvolume bosonic and superspace indices. This additional $U(1)_T$ field appears in covariant derivatives through the schematic form $D_I \sim \partial_I + A_I \delta_{\beta}$. Since there are two superspace coordinates and four supercharges, in order for the superderivatives to commute with the supercharges the connection $A_I$ must transform under the  topological Schwinger-Keldysh and KMS symmetries.

One way to think about the $U(1)_T$ field is as follows. The dynamics of the Langevin equation may be encoded in a Schwinger-Keldysh path integral~\cite{zinn2002quantum}. The ensuing action may be understood as a Schwinger-Keldysh effective action of the sort considered in our work. This action is invariant under only two of the four supercharges posited by HLR. As shown by HLR in~\cite{Haehl:2015foa}, one can render this action invariant under all four supercharges by introducing a dynamical $U(1)_T$ field.

The authors of \cite{Haehl:2015foa,Haehl:2015uoc} then hypothesized that the Grassmannian component of the field strength of $A_I$, $F_{\theta\btheta}$, spontaneously condenses with $\langle F_{\theta\btheta} \rangle= - i$. It is interesting to note that the superderivatives we constructed, c.f. \eqref{E:superDerivatives}, $D_{\theta} = \partial_{\theta} - i \btheta \delta_{\beta}$ and $D_{\btheta} = \partial_{\btheta}$, may be interpreted as having a background $U(1)_T$ connection with this same field strength. We believe that the similarity between the superderivatives is not accidental. In a future publication we will argue that one may add an external source, $A_I(\sigma)$, which couples to the current generated by the transformation $\delta_{\beta}$  (see \eqref{E:comsimple}). After such a procedure, one finds that $A_I(\sigma)$ effectively acts as a connection through the form espoused in~\cite{Haehl:2015foa,Haehl:2015uoc}.

From a structural perspective the action we propose is the sum of two terms (see \eqref{E:superdecomposition}). The first uses an $r$/$a$ superfield $\super{O}$, and the second uses an $\tilde{r}/\tilde{a}$ superfield $\wt{\super{O}}$. We refer to the second term (which seems to be absent in HLR's construction~\cite{Haehl:2015foa,Haehl:2015uoc}) as the KMS partner of the first. The partner action ensures that the total action respects the full KMS symmetry. This, in turn, guarantees that the fluctuation-dissipation relation and its non-linear generalizations~\cite{Wang:1998wg,Crossley:2015evo} hold.

A final difference between our work and that of~\cite{Haehl:2015foa,Haehl:2015uoc} concerns worldvolume symmetries. In section~\ref{S:dof} we have argued that the dynamical fields are the mappings $\super{X}^{\mu}$ and the phases $\super{C}$. Even though we have a superspace, we only imposed worldvolume reparamaterization  and $U(1)$ symmetries on the effective action. In contrast, the authors of~\cite{Haehl:2015foa,Haehl:2015uoc} employ a dynamical superembedding $\super{X}^I$ in addition to $\super{C}$, and impose superreparameterization and super-$U(1)$ invariances. We note in passing that in our probe analysis, one can show that the terms allowed by $U(1)$-invariance can be upgraded to be super-$U(1)$-invariant.

While the notation we use is very similar to \cite{Haehl:2015foa,Haehl:2015uoc}, the construction we use is, in practice, very reminiscent of that of \cite{Crossley:2015evo}. In fact we have checked that, in the probe limit, our action largely agrees with that of CGL~\cite{Crossley:2015evo}. 
As far as we could tell, the main difference between our actions has to do with the discrete symmetries imposed on the generating function. While we demanded $\CPT$, CGL demanded $\mathcal{P}\mathcal{T}$. This seems to generate a slight mismatch in the parity of observables under $\mu \to -\mu$.
The perceptive reader might worry that in addition, CGL advocated for a partial diffeomorphism and gauge invariance on the worldvolume while we uphold a full diffeomorphic and gauge invariant theory. Recall however that the initial state parameters $\beta^i$ and $\Lambda_{\beta}$ are fixed data. The residual transformations that leave $\beta^i$ and $\Lambda_{\beta}$ invariant are exactly those used by CGL, implying that the symmetries of both theories are the same.

Our main contributions to the CGL~\cite{Crossley:2015evo} construction include amalgamating fields into superfields, providing an a priori argument for the existence of a topological KMS symmetry, and imposing the full  KMS symmetry via the introduction of tilde'd superfields. Most importantly, we have redefined the superfields using $\Rr(i\delta_{\beta})$ and $\Aa(i\delta_{\beta})$ so that the associated transformations and superderivatives satisfy the Leibniz rule (even beyond the $\hbar \to 0$ limit discussed in \cite{Crossley:2015evo,Glorioso:2016gsa}). 

To summarize,  in this work we have constructed a superspace formalism of dissipative hydrodynamics with {a} notation similar to {that of} \cite{Haehl:2015foa,Haehl:2015uoc}, but formulation closer in spirit to {that of} ~\cite{Crossley:2015evo}. We have demonstrated that the hydrodynamic constitutive relations and the fluctuation-dissipation relations are compatible with our formalism.  This is, of course, a minimal requirement of  {a Schwinger-Keldysh} effective action for hydrodynamics. With a full fledged effective action one can do much more.  

\subsection{Outlook}

We now turn our attention to loose ends and open questions which are suggested by our work. One such loose end is the treatment of hydrodynamic frame transformations~\cite{LL6} in the effective action. In hydrodynamics, one may redefine the hydrodynamic variables to eliminate unphysical transport coefficients, and we expect this redefinition to descend from an operation in the Schwinger-Keldysh action. While we do not know with certainty how to implement this redefinition, we have preliminary indications that it descends from supersymmetry-preserving redefinitions of the sigma model superfields.

Another loose end which we need to address is a complete effective action which goes beyond the probe limit. In section \ref{S:action} we have demonstrated that, in the probe limit, the symmetry requirements listed in section \ref{S:Ingredients} lead to appropriate constitutive relations and fluctuation dissipation relations.  As discussed in section \ref{S:dof} all our arguments easily go through when dealing with the full hydrodynamic theory, but restricted to quadratic order in the $a$-type fields. Preliminary investigations suggest that it is straightforward to correct the action perturbatively in $a$-type fields.

An additional prospect for the future involves a finer study of transport. There are qualitatively different types of transport that may be realized in the most general hydrodynamic setting, as classified in~\cite{Haehl:2015pja}. It remains to be demonstrated that all classes of transport may be realized through effective actions of the sort studied in this paper (or, for that matter, by those used by CGL \cite{Crossley:2015evo} and HLR \cite{Haehl:2015foa,Haehl:2015uoc}). In particular, it would be interesting to determine the modifications to the action which are necessary to match `t Hooft anomalies, which would account for anomaly-induced transport~\cite{Son:2009tf} (see also~\cite{Jensen:2013kka,Jensen:2013rga}).
	 
Another question concerns the status of our effective actions as full-fledged quantum theories. One point of concern in this respect is that the sigma model may have zero modes, and when this occurs we expect that they must be quotiented out. Our motivation for this is a bit oblique. We did not emphasize it before, but the low-energy description of the Sachdev-Ye-Kitaev (SYK) models at large $N$ and low temperature~\cite{Maldacena:2016hyu}, or the theories dual to dilaton gravity on a nearly AdS$_2$ spacetime~\cite{Jensen:2016pah,Maldacena:2016upp,Engelsoy:2016xyb}, is a $0+1$d sigma model of the type discussed in this paper. In the SYK models and two-dimensional gravity, one must quotient out this sigma model by a $SL(2;\mathbb{R})$ symmetry which acts as $1d$ conformal transformations on the worldvolume~\cite{Maldacena:2016hyu}.

The main reason for focusing our attention so far to hydrodynamics is practical. Much is known of hydrodynamics on phenomological grounds, and so it offers a useful testing ground to nail down the correct principles for Schwinger-Keldysh effective field theory. One of the most unusual features of these models is the existence of an entropy current. As we alluded to above, one may use the effective action to argue for the existence of an entropy current with a positive semi-definite divergence (see  \cite{Glorioso:2016gsa}). We will also explore the entropy current in a future publication, tying it to the $U(1)_T$ symmetry proposed by \cite{Haehl:2015foa,Haehl:2015uoc}.

On a more fundamental level, it is important to understand whether there is a Schwinger-Keldysh no-ghost theorem. In this work we have been deliberately vague regarding the implementation of a ghost number symmetry. We have mentioned that a ghost number symmetry will forbid certain terms in the effective action. One way to check whether ghosts have been correctly incorporated would be to match our action to a theory where one may reliably compute quantum corrections to hydrodynamic correlators. A prime candidate for such a theory is the AdS/CFT correspondence. We hope to report on this issue in the near future.

A formal aspect of this work which we have considered in detail is the KMS symmetry. We have first argued that the full KMS symmetry of the Schwinger-Keldysh generating function implies the existence of a topological KMS symmetry. In this sense the topological KMS symmetry emerges once the full KMS symmetry is implemented. In generating the effective action we have found it useful to first implement the topological KMS symmetry and only then impose the full KMS symmetry. Clearly, with some work, one should be able to implement the full KMS symmetry in one go and obtain the topological sector as a result.

Once the correct principles for constructing the Schwinger-Keldysh effective theory are  known, one can then use the lessons learned to tackle systems where much less is known. An obvious place to start is with the generalizations of the Schwinger-Keldysh partition function which encode out-of-time-ordered correlators, as emphasized by~\cite{Haehl:2016pec,Haehl:2017qfl} (see also~\cite{Aleiner:2016eni}). These have been the subject of intense study from the point of view of diagnosing early-time chaotic growth in many-body systems (see e.g.~\cite{Maldacena:2015waa}). In particular, in generalizations of the SYK models~\cite{Gu:2016oyy,Davison:2016ngz} and in holography~\cite{Blake:2016sud,Blake:2016jnn} there is a curious relation between the ``butterfly velocity'' appearing in the exponential growth of out-of-time-ordered four-point functions, which determines the speed at which the chaotic growth propagates, and the underlying diffusion constants. Perhaps this relation follows from the symmetries of an effective theory on the four-fold contour, which generalizes the Schwinger-Keldysh hydrodynamics of our work.

Finally, it is an open problem to realize the Schwinger-Keldysh effective descriptions in  this work from the AdS/CFT correspondence. While some crucial first steps toward this goal were made in \cite{deBoer:2015ija,Crossley:2015tka}, the matter is far from settled. For example, in our effective actions there are ghost partners of the sigma model fields, but there is (as of yet) no sign of these ghosts in a dual gravitational description. It is conceivable that a proper treatment of the two-sided black hole, within the Schwinger-Keldysh formalism, may shed light on this puzzle. This, in turn, may shed light on what ought to be meant by the ER=EPR correspondence~\cite{Maldacena:2013xja}. 

\acknowledgments

We would like to thank P.~Glorioso, F.~Haehl, A. Karch, P.~Kovtun, H.~Liu, R.~Loganayagam, R.~Marjieh, R.~Myers, M.~Rangamani, S.~Razamat, A.~Ritz, and D.~Teaney for many enlightening conversations. The work of NPF was supported in part by the Israel Science Foundation under grant 504/13 and in part at the Technion by a fellowship from the Lady Davis Foundation. The work of AY was supported in part by the Israeli Science Foundation under an ISF-UGC grant 630/14 and an ISF excellence center grant 1989/14.

\bibliographystyle{JHEP}
\bibliography{Formalismbib}
\end{document}